\PassOptionsToPackage{dvipsnames}{xcolor}
\PassOptionsToPackage{colorlinks}{hyperref}

\documentclass[acmsmall,screen,nonacm]{acmart}

\AtBeginDocument{%
  }

\acmJournal{JACM}
\acmVolume{XX}
\acmNumber{X}
\acmArticle{}
\acmMonth{6}

\setcopyright{none}
\renewcommand\footnotetextcopyrightpermission[1]{}

\acmSubmissionID{23}

\usepackage[subtle]{savetrees}

\makeatletter
\renewcommand\paragraph{\@startsection{paragraph}{4}{0pt}%
  {-.5\baselineskip \@plus -2\p@ \@minus -.2\p@}%
  {-3.5\p@}%
  {\ACM@NRadjust{\bf\@adddotafter}}}
\makeatother

\usepackage{microtype}
\microtypecontext{spacing=nonfrench} 

\usepackage[scale=0.8]{cascadia-code}
\usepackage[T1]{fontenc}
\usepackage[dvipsnames]{xcolor}

\usepackage{xspace}
\usepackage{outlines}
\usepackage{enumitem}      
\usepackage{soul}           
\usepackage[english]{babel}
\usepackage[rightcaption]{sidecap}

\usepackage{amsthm}  

\usepackage{tabularx}
\usepackage{colortbl}
\usepackage{multicol}

\usepackage{tikz}

\usepackage{subcaption}
\usepackage[labelfont={bf, small}, font={small}]{caption}
\numberwithin{equation}{section}

\newcommand{\notes}[1]{}
\newcommand{\anup}[1]{\notes{{\color{red}AA: {#1}\xspace{}}}}
\newcommand{\va}[1]{\notes{{\color{Green}VA: {#1}\xspace{}}}}
\newcommand{\srini}[1]{\notes{{\color{purple}SS: {#1}\xspace{}}}}

\newcommand{\cut}[1]{} 

\newcommand{\heading}[1]{\smallskip\noindent\textbf{#1.}}
\newcommand{\subheading}[1]{\emph{#1.}}

\newcommand{\myparagraph}[1]{\heading{#1}}
\newcommand{\mysubparagraph}[1]{\subheading{#1}}

\newcommand{\eg}{e.g.\xspace}
\newcommand{\Eg}{E.g.\xspace}
\newcommand{\etc}{etc.\@\xspace}
\newcommand{\cf}{{cf.}\xspace}
\newcommand{\ie}{i.e.\xspace}
\newcommand{\Ie}{I.e.\xspace}

\newenvironment{packeditemize}{
\begin{itemize}[topsep=0pt, noitemsep, leftmargin=*]}
{\end{itemize}}

\newenvironment{packedenumerate}[1][(\arabic{*})]{
\begin{enumerate}[topsep=0pt, noitemsep, leftmargin=*, label=#1]}
{\end{enumerate}}

\newtheoremstyle{mytheoremstyle} 
    {\topsep}                    
    {\topsep}                    
    {\itshape}                   
    {}                           
    {\scshape}                   
    {.}                          
    {.5em}                       
    {}  
\theoremstyle{mytheoremstyle}

\newtheorem*{theorem*}{Theorem}  

\theoremstyle{mytheoremstyle}

\theoremstyle{plain}

\theoremstyle{remark}

\DeclareRobustCommand{\hlviolet}[1]{{\sethlcolor{violet!10}\hl{#1}}}
\DeclareRobustCommand{\hlgreen}[1]{{\sethlcolor{green!10}\hl{#1}}}
\DeclareRobustCommand{\hlblue}[1]{{\sethlcolor{blue!10}\hl{#1}}}

\definecolor{TableGray}{rgb}{0.88,0.88,0.88}
\definecolor{TextGray}{rgb}{0.37647,0.37647,0.37647}

\newcommand{\ns}[1]{\ensuremath{#1}\xspace}       

\newcommand{\bw}[1][]{\ns{C_{#1}}}

\newcommand{\rtprop}[1][]{\ns{R_{#1}}}

\newcommand{\delay}{\ns{\texttt{delay}}}

\newcommand{\hops}[1][]{\ns{\texttt{hops}}}
\newcommand{\rtt}[1][]{\ns{\texttt{RTT}_\texttt{#1}}}
\newcommand{\cwnd}[1][]{\ns{\texttt{cwnd}_\texttt{#1}}}
\newcommand{\rate}[1][]{\ns{\texttt{rate}_\texttt{#1}}}

\newcommand{\utility}{\ns{U}}              
\newcommand{\x}[1][]{\ns{r_{#1}}}            
\newcommand{\s}[1][]{\ns{s_{#1}}}            
\newcommand{\f}[1][]{\ns{f_{#1}}}            

\newcommand{\Sig}[1][]{\ns{S_{#1}}}            
\newcommand{\Smin}{\ns{\Sig[\min]}}
\newcommand{\Smax}{\ns{\Sig[\max]}}
\newcommand{\Cmin}{\ns{\bw[\min]}}
\newcommand{\Cmax}{\ns{\bw[\max]}}

\newcommand{\func}{\ns{\texttt{func}}}
\newcommand{\sfunc}{\ns{\texttt{f}}}
\newcommand{\agg}[1][]{\ns{\texttt{fold}_{#1}}}
\newcommand{\grow}{\ns{\texttt{growth}}}
\newcommand{\ratio}{\ns{\texttt{ratio}}}
\newcommand{\ds}{\ns{\delta\s}}

\newcommand{\ef}{\ns{\epsilon_{\x}}}
\newcommand{\rstar}{\ns{\x^\star}}
\newcommand{\sstar}{\ns{\s^\star}}
\newcommand{\bwstar}{\ns{\bw^\star}}

\newcommand{\ratiostar}{\ns{\texttt{ratio}^\star}}

\newcommand{\cwndupdateavg}{\ns{\alpha}}

\newcommand{\tcwnd}[1][]{\ns{\texttt{target\_cwnd}_\texttt{#1}}}

\begin{document}

\setlength{\abovedisplayskip}{3pt}
\setlength{\belowdisplayskip}{4pt}
\setlength{\abovedisplayshortskip}{0pt}  
\setlength{\belowdisplayshortskip}{0pt}



\allowdisplaybreaks

\title{Contracts: A unified lens on congestion control robustness, fairness, congestion, and generality}

\author{Anup Agarwal}
\email{anupa@cs.cmu.edu}
\affiliation{%
  \institution{Carnegie Mellon University}
  \city{Pittsburgh}
  \state{Pennsylvania}
  \country{USA}
}
\author{Venkat Arun}
\email{venkat@utexas.edu}
\affiliation{%
  \institution{University of Texas}
  \city{Austin}
  \state{Texas}
  \country{USA}
}
\author{Srinivasan Seshan}
\email{srini@cs.cmu.edu}
\affiliation{%
  \institution{Carnegie Mellon University}
  \city{Pittsburgh}
  \state{Pennsylvania}
  \country{USA}
}

\renewcommand{\shortauthors}{Agarwal et al.}

\begin{abstract}

Congestion control algorithms (CCAs) operate in partially observable
environments, lacking direct visibility into link capacities, or competing
flows. To ensure fair sharing of network resources, CCAs communicate their fair
share through observable signals. For instance, Reno's fair share is encoded as
$\propto 1/\sqrt{\texttt{loss rate}}$. We call such communication mechanisms
\emph{contracts}. We show that the design choice of contracts fixes key
steady-state performance metrics, including robustness to errors in congestion
signals, fairness, amount of congestion (e.g., delay, loss), and generality
(e.g., range of supported link rates). This results in fundamental tradeoffs
between these metrics. Using properties of contracts we also identify design
pitfalls that lead to starvation (extreme unfairness). We argue that CCA design
and analysis should start with contracts to conscientiously pick tradeoffs and
avoid pitfalls. We empirically validate our findings and discuss their
implications on CCA design and network measurement.

\anup{Remove measurement?}



\end{abstract}

%
%
%

\maketitle

\section{Introduction}


Congestion control algorithms (CCAs) play a critical role in ensuring fair and
efficient bandwidth allocation. Despite this, reasoning about their fairness
has been ad hoc. Traditionally, CCAs were designed to avoid congestion
collapse, and their fairness properties were analyzed
retrospectively~\cite{fairness-chiu-jain,num-srikant,num-low,num-kelly}. With
the rise of interactive applications and short remote-procedure calls (RPCs),
newer CCAs prioritized (local) performance metrics like latency and convergence
time, where traditional CCAs performed poorly. However, this ``local''
perspective often led to undesirable fairness. For instance,
TIMELY~\cite{timely} has been shown to have infinite fixed points, causing
unfair rate allocations~\cite{ecn-vs-delay}.


The result has been that a CCA that meets key efficiency
(generality), latency (congestion), fairness, and robustness objectives remains elusive. 
Not only are real-world networks complicated and hard to
predict, but empirical and theoretical evidence suggests that all objectives
are not simultaneously achievable~\cite{starvation, ccmatic, ecn-vs-delay,
throughput-vs-fairness, faircloud,num-low-book}. Thus, any CCA design must make
difficult tradeoffs. 
Any systematization that reveals and navigates these tradeoffs can guide better
CCA design.



\myparagraph{Contributions} We contribute to this effort by (C1) building an
abstraction, that we call {\em contracts}, to formalize the mechanisms CCAs use to coordinate
fairness (\autoref{sec:motivation}, \autoref{sec:contracts}). This led us to
(C2) discover and analytically quantify previously unknown tradeoffs across a
broad class of CCAs (\autoref{sec:metrics-tradeoffs}), and (C3) identify
previously unknown scenarios and (avoidable) design mistakes that cause severe
performance degradation or suboptimal performance in a wide range of CCAs,
including, BBR~\cite{bbr}, Vegas~\cite{vegas}, ICC~\cite{icc},
Swift~\cite{swift} (\autoref{sec:motivation}, \autoref{sec:pitfalls}). Our
methodology applies to analytically designed CCAs, 
black-box implementations, such as Tao (Remy)~\cite{tao, remy}, and learned CCAs,
such as PCC~\cite{pcc} (online-learning) and Astraea~\cite{astraea}
(reinforcement-learning). Using contracts and our learnings, we (C4) build
blueprints for CCA design and analysis that make the tradeoffs and mistakes
intuitive and actionable (\autoref{sec:motivation}). We hope this will help
avoid performance mistakes and enable conscientious selection of Pareto-optimal
tradeoffs in future designs. (C5) We validate our findings using simulation and
emulation, finding a near-perfect match between analysis and measurements
(\autoref{sec:empirical}).


\myparagraph{Contracts} CCAs operate in partially observable
environments---unaware of the network topology, link capacities, routing, or
number of competing flows. To achieve fairness, CCAs encode fair shares into
observable network signals. For instance, Reno uses loss rate to coordinate
fair shares, where each endpoint transmits at an average rate $\propto
1/\sqrt{\texttt{loss rate}}$. We call such communication mechanisms
\emph{contracts}. To our knowledge, all existing fair CCAs have a
contract in their design, either implicitly (\eg, Reno~\cite{mathis-model},
Vegas~\cite{duality-vegas}), or explicitly (\eg, TFRC~\cite{tfrc},
Swift~\cite{swift}, Poseidon~\cite{poseidon}).



CCA contracts differ in the signals they use and their shape (\eg, steeper vs
gradual). For example, delay-based CCAs encode fair share as: ``$1 /
\texttt{delay}$'' \cite{vegas,fasttcp,copa}, ``$1 / \texttt{delay}^2$''
\cite{swift}, or ``$e^{-\texttt{delay}}$'' \cite{starvation}. Throughout the
paper, we use delay to mean a queuing delay estimate (\eg $\rtt -
\min \rtt$).
Contracts effectively parameterize the space of CCAs without fully specifying a
CCA. For instance, the entire family of TCP-friendly CCAs~\cite{reno, newreno,
tfrc, binomial-infocom} ``follow'' Reno's contract (\ie send at rates $\propto
1 / \sqrt{\texttt{loss rate}}$), but they differ in other aspects, such as
stability and convergence time properties.


\myparagraph{Tradeoffs} This variety in contracts raises the question: ``What
are the tradeoffs between different contracts?'' In exploring this question, we
discovered a surprising result. The contract---a design aspect that may not
have been chosen explicitly---fully determines four key congestion control
metrics: (1) \emph{robustness} to noise in congestion signals, (2)
\emph{fairness} in a multi-bottleneck network, (3) amount of \emph{congestion}
(\eg, delay, loss), and (4) \emph{generality} (\eg, the range of link rates the
CCA supports). Robustness and fairness are better with gradual contracts, while
congestion and generality are better with steeper contracts
(\autoref{sec:metrics-tradeoffs}). Thus, we cannot have the best in all four
metrics.

Note that tolerating more congestion allows supporting a larger bandwidth
range. In that sense congestion and generality are also at odds. Further,
fairness definitions (\autoref{sec:metrics-tradeoffs}) subsume link utilization
and throughput. The tradeoff between fairness and throughput has been
well-studied~\cite{alpha-fairness}, so we do not re-discuss it but acknowledge
that the contract choice also fixes this tradeoff.


Similar tradeoffs have been explored before. \cite{starvation} discusses a
special case where full generality precludes simultaneous robustness and
bounded variation in congestion. We find that their proposed workarounds cause
unfairness due to other reasons (\autoref{sec:metrics-tradeoffs},
\autoref{sec:pitfalls}). \cite{ecn-vs-delay} shows that delay-based CCAs cannot
simultaneously ensure fixed delays and fairness. That is, if a delay-based CCA
maps the same delay value to multiple rates, it does not have a contract, and
cannot be fair.
The network utility maximization (NUM) literature~\cite{num-kelly, num-srikant,
num-low, duality-vegas} studies a subset of these tradeoffs for individual
CCAs. To our knowledge, we are the first to uncover and generalize these
tradeoffs to many CCAs.

\myparagraph{Mistakes} We use contracts to identify a variety of avoidable
design mistakes (\autoref{sec:motivation}, \autoref{sec:pitfalls}). CCAs whose
contracts have extreme shapes (\eg, logarithmic, exponential), shifts, clamps,
or intercepts can starve flows. We observe this in BBR, ICC, and Astraea. CCAs with
an explicit contract (\eg, Swift) do not need AIMD updates to reach fairness.
Instead, MIMD updates allow exponentially fast convergence to both fairness and
efficiency (\autoref{sec:canonical}). Finally, we show that having fixed
end-to-end thresholds, \eg, AIMD on delay \cite{1rma, smartt, strack}, causes
starvation on topologies with multiple bottlenecks.


\myparagraph{Blueprints} We argue for a contracts-first approach to CCA design
and analysis. Contracts are intuitive and actionable. The tradeoffs are obvious
from the contract choice (after a few to no algebraic steps) and contracts
align with how our community designs CCAs: a variety of CCAs start with a
target rate equation and build dynamics around it, \eg, Swift~\cite{swift},
TFRC~\cite{tfrc}.

Compared to other design patterns, contracts are better suited to prevent
mistakes. Many methodologies start from a utility function: local utilities
(\eg, ``$\log(\texttt{tput}) - \log(\delay)$'' in Copa~\cite{copa}), global
utilities (\eg, NUM/Remy~\cite{remy}), or reward functions in
RL~\cite{astraea}. These hide tradeoffs giving an illusion of unilaterally
optimizing latency or throughput when these are intertwined with other metrics
like fairness, robustness, and generality (\autoref{sec:metrics-tradeoffs}).
Consequently, many designs pick sub-optimal tradeoffs as inadvertent artifacts
of other design decisions (\autoref{sec:contracts},
\autoref{sec:metrics-tradeoffs}, \autoref{sec:pitfalls}). Contracts, on the
other hand, force the designer to choose between uncomfortable tradeoffs and
pick Pareto-optimal points conscientiously.



We end by discussing the limitations of contracts and how we may expand their
uses (\eg, reasoning about inter-CCA fairness), and potential ways to work
around the tradeoffs (\autoref{sec:limitations}, \autoref{sec:discussion}).

\section{Motivation (Why do we need contracts?)}
\label{sec:motivation}



\begin{figure*}[t]
    \centering
    \begin{minipage}[b]{0.32\textwidth}
        \includegraphics[width=\linewidth]{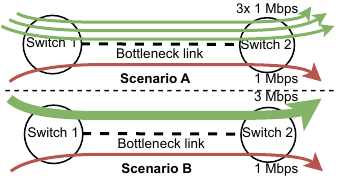}
        \caption{\label{fig:assumptions-other-flows} To distinguish the 
        scenarios, the red (lower) flow must coordinate with green
        (upper) flows.}
    \end{minipage}\hfill
    \begin{minipage}[b]{0.32\textwidth}
        \includegraphics[width=\linewidth]{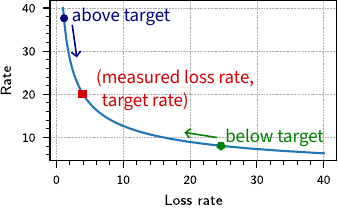}
        \caption{\label{fig:reno-contract} Reno uses loss rate to communicate fair
            shares.}
    \end{minipage}\hfill
    \begin{minipage}[b]{0.32\textwidth}
        \includegraphics[width=\linewidth]{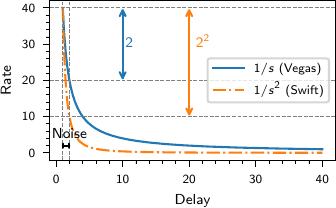}
        \caption{\label{fig:contracts:robustness} Steeper contract implies worse
            robustness to noise.}
    \end{minipage}
\end{figure*}


Since CCAs do not know how many flows they are competing against,  they need
some form of agreement between flows to determine their fair share.
For instance, consider the scenarios in \autoref{fig:assumptions-other-flows}.
If the green (upper) flow in scenario B exactly emulates the cumulative effect
of the three green flows in scenario A, then the red flow cannot tell the
difference. Its observations (timestamps of when packets are sent and
acknowledged) are identical. Yet, its fair share is different.

Contracts help flows to disambiguate between such scenarios. Since flows cannot
directly communicate with each other, all flows ``agree'' to follow an encoding
of fair share into globally observed signals. For example, Reno uses:
``$\texttt{rate} \propto 1/\sqrt{\texttt{loss rate}}$''~\cite{mathis-model,
tfrc}. While this is not explicit in the design, the emergent behavior is
that all flows react to losses such that they effectively measure the average packet loss rate and calculate a ``target
rate'' using this formula. Then they increase or decrease their actual sending
rate to move towards this target (\autoref{fig:reno-contract}). Returning to
scenario B in \autoref{fig:assumptions-other-flows}, both the red and green
flows measure the same loss rate, and hence calculate the same target rate, but
they have different sending rates. Thus, at least one of them will change their
rate until they reach a steady-state where everyone's sending rates are equal.

Mathematically, the contract forces a unique equilibrium for flows. Consider a
fluid model execution of a CCA on a dumbbell topology, with $n$ flows (\f[1],
to \f[n]), flow \f[i] having an RTprop (round-trip propagation delay) of
\rtprop[i] seconds and a link capacity of \bw packets/second. The fluid model
equations yield $n+1$ independent variables with $1$ independent equation:
$\sum_i \rate[i] = \bw$. Here, $\rate[i] = \cwnd[i]/\rtt[i]$ and $\rtt[i] =
\delay + \rtprop[i]$. \cwnd[i] and \delay are the independent variables. The
CCA through its contract (\eg, $\rate[i] = \cwnd[i]/\rtt[i] = 1/\delay$) yields
$n$ additional equations to ensure a unique solution to this system.


To our knowledge, all existing CCAs use a similar method of coordination. The
signals used and contract shapes vary, but the underlying concept remains the
same. \Eg,
DCTCP, DCQCN, and MPRDMA use average ECN marking rate~\cite{dctcp, dcqcn,
mprdma, ecn-vs-delay}; Poseidon uses the maximum per-hop delay~\cite{poseidon};
AIMD on delay uses bytes (or time) between high delay (\autoref{sec:pitfalls}).
Even CCAs generated through machine-learning (\eg, Astraea~\cite{astraea})
implicitly learn to use contracts (\autoref{sec:compute}).


Due to the coupling between rate and congestion, a CCA's choice of contract induces
a variety of tradeoffs, \eg, halving the link capacity quadruples the
steady-state loss rate for Reno (\autoref{sec:metrics-tradeoffs}).

\anup{Elaborate more on this? as we talk a lot about tradeoffs and implications
on CCA design before describing the details of the tradeoffs.}

\subsection{Isn't this obvious? Why are we writing this paper?}


Contracts share mathematical foundations with NUM (network utility
maximization)~\cite{num-kelly, num-srikant, num-low, duality-vegas}
(\autoref{sec:contracts}) and some of our results may seem like simple extensions of
existing results.
Despite this, many CCA designs (including recent ones) repeatedly commit
avoidable mistakes resulting in poor performance.
We detail some of these mistakes below with others in \autoref{sec:pitfalls}.
To our knowledge, outside of mistake \#1, other mistakes (including
\autoref{sec:pitfalls}) have not been documented before. Contracts helped us
discover these issues, and make it easier to systematically avoid them in
future designs.


These mistakes stem from several anti-patterns, including, treating CCA design
as a mere reaction to congestion (without analytically understanding their
consequences) or blindly using AIMD in hopes of ensuring fairness. The most
prominent anti-pattern is myopically optimizing latency without reasoning about
fairness. This manifests in various ways, including, setting constant delay
targets in hand-designed CCAs (\autoref{sec:pitfalls}) or in reward functions
of RL-based designs (\autoref{sec:contracts}, \autoref{sec:pitfalls}). However,
latency is intertwined with other metrics and cannot be optimized unilaterally
(\autoref{sec:metrics-tradeoffs}, \autoref{sec:pitfalls}). Note that this is
different from the tradeoff between latency and throughput, which stems from
variations in link capacity as opposed to the coordination mechanisms used by
CCAs.


\anup{Another anti-pattern: artificially restricting the CCA design space with
justification as ease in hardware implementation.}






\myparagraph{Mistake \#1: Not having a contract} CCAs like TIMELY~\cite{timely}
do not have a contract, \ie, there is no unique mapping between steady-state
rate and congestion signals to force a unique equilibrium. As shown
in~\cite{ecn-vs-delay}, TIMELY admits infinitely many solutions to the
steady-state equations described above. Each solution corresponds to a
different allocation of flow rates (provided they sum to link capacity), with a
potentially arbitrarily large ratio of flow rates (arbitrary unfairness).

\myparagraph{Mistake \#2: Picking extreme contracts} Recent proposals like
ICC~\cite{icc}, Astraea~\cite{astraea}, and the exponential CCA
from~\cite{starvation}, optimize for a narrow subset of performance metrics,
yielding contracts that are good for some metrics but extremely bad for others.
All three CCAs cause extreme unfairness in multi-bottleneck topologies
(\autoref{sec:metrics-tradeoffs}). ICC and Astraea are also extremely
susceptible to network jitter (\autoref{sec:metrics-tradeoffs}).

\myparagraph{Mistake \#3: Conflating AIMD with contracts, and picking
sub-optimal dynamics} Swift~\cite{swift} uses AIMD to update cwnd despite
having an explicit contract. AIMD is not necessary for fairness and
unnecessarily increases convergence time. With an explicit contract, we can
perform MIMD updates to reach fairness exponentially fast while maintaining a
stable control loop (\autoref{sec:canonical}). For instance, in
\autoref{fig:reno-contract}, all flows measure the same loss rate and compute
the same target rate. Flows can update their current rate to move toward the
target using any increment choice (additive or multiplicative). Fairness is
ensured despite MIMD updates because flows stop changing their rate \emph{if
and only if} all flow rates are equal to the target rate (and hence to each
other).


Likewise, PowerTCP~\cite{powertcp} claims that RTT-gradient based CCAs
(``current-based'' in~\cite{powertcp}) are more reactive and precise than
RTT/delay/loss/ECN-based CCAs (``voltage-based'' in~\cite{powertcp}). We argue
that reactivity is orthogonal to the choice of congestion signal. We can have
exponentially fast convergence to both fairness and efficiency even when using
``voltage-based'' control (\autoref{sec:canonical}).

Finally, works like Poseidon~\cite{poseidon} artificially distinguish AIMD
control and ``target scaling'' (contracts) even though they are mathematically
equivalent. AIMD \emph{implicitly} creates a contract (or scaling/mapping
between rate and congestion), while ``target scaling'' is an explicit contract.

\subsection{Avoiding the mistakes using a contract-first blueprint of CCA
design/analysis}

While there is no exhaustive list of mistakes, we hope that by following a
contract-first blueprint, many common mistakes can be identified and avoided.

\myparagraph{Design blueprint} To design a new CCA, first, (D1) pick a
contract. In \autoref{sec:contracts}, we define contracts as a function.
Picking a contract involves deciding its (D1.1) input (\eg, delay, loss, ECN,
\etc) and output (\eg, rate, cwnd, fraction of link, \etc), (D1.2) shape (\eg,
linear, square-root, exponential, \etc), and (D1.3) parameters (\eg, scale,
shift, clamps, \etc). \autoref{sec:metrics-tradeoffs}, \autoref{sec:pitfalls}
and \autoref{sec:workarounds} give guidance on how these choices affect
steady-state performance. Then, (D2) implement dynamics to follow the contract.
\autoref{sec:canonical} gives guidance on how the dynamics impact convergence
time/stability along with other design considerations. Note that unless
congestion control is solved, our list of considerations is necessarily
incomplete. \srini{not clear what solve means here} \anup{A complete list would
imply solving CC.}


\myparagraph{Analysis blueprint} Similarly, for analyzing an existing CCA, one
should: (A1) compute its contract (\autoref{sec:compute}), (A2) see where the
contract lies in the tradeoff space (\autoref{sec:metrics-tradeoffs}), (A3)
identify any obvious issues due to shifts/clamps in the contract
(\autoref{sec:pitfalls}), and (A4) compare dynamics with those in
\autoref{sec:canonical}.

\autoref{fig:assumptions-other-flows} shows why some form of agreement between
flows is necessary to achieve fairness with end-to-end CCAs. While we do not
formally prove this, we believe this agreement can always be represented as a
contract function, and consequently, the tradeoffs induced by contracts are
fundamental. The only way to work around the tradeoffs is to change the
input/output in the contract to decouple performance metrics from the contract
(\autoref{sec:workarounds}). We believe that other efforts for improving
steady-state performance are futile and will likely lead to reinventing a CCA
already covered in the design/tradeoff space in~\autoref{sec:metrics-tradeoffs}.
\srini{perhaps shrink this a bit -- seems repetitive with earlier}

\va{"agreement can be written as function" this is a very strong statement.
  E.g. even for FRCC, you need to do some mental gymnastics to see the contract
  as a function. And with that level of generality in the inputs/outputs of
  that contract function, the statement becomes a little vacuous. We should
either prove it (which shouldn't be too hard) or remove this claim.} \anup{Move
this para to workarounds section?~\autoref{sec:workarounds}? I weakly refer to
it in tradeoffs.}



\section{Contracts (Definition)}
\label{sec:contracts}


The contract of a CCA is a function of the form $\texttt{avg sending rate}$ $ =
\func(\texttt{aggregate statistic})$ that describes the CCA's steady-state
behavior (\eg, at time infinity), when competing with itself on a dumbbell or
parking lot topology (\autoref{fig:parking_lot}). We use these topologies to
define the performance metrics \autoref{sec:metrics-tradeoffs}, so we only need
the contracts to hold on these topologies. As a shorthand, we use ``$\x =
\func(\s)$'', or ``$\x = \sfunc(\s)$''. Here, $\s$ is an aggregate statistic
derived from the CCA's observations: the time series of sending and ACK
sequence numbers, and any explicit signals (\eg, ECN~\cite{ecn}). Statistics
like delay, loss, and ACK rate can be derived from the time series.

\srini{a bit wordy} \anup{"describe" has no mathematical meaning. But we do not
have a more precise definition that is correct. We can say, the CCA behavior is
always a subset of the contract. And tighter contracts are more meaningful.}



We only consider strictly decreasing contracts, with closed intervals as their
domain and range, ensuring they are continuous and invertible. All contracts
that we are aware of are decreasing. The statistic typically measures
congestion. An increasing contract suggests increasing the rate with increasing
congestion. This further increases congestion, creating a positive feedback
loop.



To enable reasoning about a variety of CCAs---which encode information in
different signal properties, such as magnitude (\eg, delay) or statistical
moments (\eg, loss rate)---our contract definition allows flexibility in
picking \s. This freedom needs to be exercised judiciously. For instance, for
Vegas, it is more natural to pick $\s = \delay$ with $\func(\s) = 1/\s$, as
opposed to $\s = \delay^2$ with $\func(\s) = 1/\sqrt{\s}$. Both representations
have identical performance. However, since we state our results in terms of \s,
interpreting our results will introduce an additional step of translating \s to
the physical quantity of delay under the latter representation.

\srini{I assume that s need to be a single statistic -- perhaps need to clarify
this?}\anup{We describe that in computing contracts part currently.}




\myparagraph{Rate- vs cwnd-based contracts} A contract can also be written in
terms of rate or cwnd, and using $\rate = \cwnd/\rtt$ to convert between the
two forms. Independent of the form, the CCA can be implemented using either
rate or cwnd (\autoref{sec:canonical}). Without loss of generality, we
consider rate-based contracts.

\myparagraph{RTT or RTprop bias} Many CCAs have an RTT or RTprop bias, \ie,
flows with different RTTs get different rates, \eg, Reno allocates more rate to
flows with lower RTT. This bias shows up as a factor in the contract, \eg,
$\rate \propto 1/(\rtt \sqrt{\texttt{loss\_rate}})$. Unlike the tradeoff
induced by the shape of contracts, this bias is not fundamental and can be
easily removed (\autoref{sec:compute}). So we omit such factors.

\myparagraph{Contracts in NUM} The concept of contracts can also be described
using NUM~\cite{num-kelly, num-srikant, num-low, num-low-book}. A contract maps
the aggregate statistic (congestion measure or price) to rate, same as the
demand function (target rate) for a given price~\cite{num-kelly}. The inverse
of a contract represents the link price (\eg, target delay) for a given load
(\eg, link's ingress rate)~\cite{num-kelly}. Consequently, the utility function
that the CCA optimizes is (derived in~\cite{num-low}): $\utility(\x) = \int
\func^{-1}(\x) \, d\x~\refstepcounter{equation}
(\theequation)\label{eq:utility}$. Here, \utility is the utility derived from a
rate of \x.
\autoref{eq:utility} only holds for statistics that add up over links (\eg,
delay). For other statistics (\eg, max per-hop delay), the utility is
different~\cite{poseidon}.
\subsection{Computing contracts}
\label{sec:compute}

\begin{figure*}[t]
    \centering
    \begin{minipage}[b]{0.32\linewidth}
        \includegraphics[]{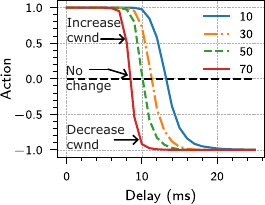}
        \caption{\label{fig:astraea} Astraea's state to action mapping (adapted from~\cite{astraea}).}
    \end{minipage}
    \hfill
    \begin{minipage}[b]{0.32\linewidth}
        \includegraphics[]{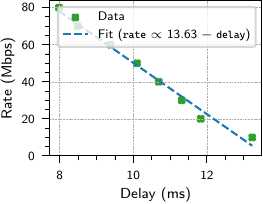}
        \caption{\label{fig:astraea-fit} Curve fitting analytical fixed-points for Astraea's contract.}
    \end{minipage}\hfill
    \begin{minipage}[b]{0.32\linewidth}
        \includegraphics[]{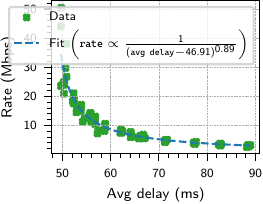}
        \caption{\label{fig:pantheon-fit} Curve fitting empirical fixed-points for BBR's contract.}
    \end{minipage}\hfill
\end{figure*}

We compute the contract of a CCA by analyzing its steady-state behavior. We run
it analytically or empirically on a variety of dumbbell and parking lot
configurations (with different capacities, RTprops, buffer sizes, and flow
counts) and collect the set of steady-state observations. We then group this
set based on an aggregate statistic, such that observations with the same
statistic value have the same rate. These equivalence classes define the CCA's
contract function.

Often, this reduces to program analysis of the CCA's code (see Vegas's example
below) or fluid model analysis (\cite{ecn-vs-delay} analyzes
DCQCN~\cite{dcqcn}). In other cases, if the aggregate statistic is
multi-dimensional and/or the contract is a compound function, the process may
require understanding CCA's internals. For instance, Copa emulates Reno on
detecting competing loss-based flows~\cite{copa}. When the loss rate is zero,
Copa follows the delay-based contract versus Reno's contract otherwise. For
simplicity, we focus on homogeneous settings where flows use the same CCA and
run in a single mode, \eg, we disable mode-switching in Copa. We expect the
different modes to follow the tradeoffs according to the contract followed in
the mode.

\anup{In such cases, it is also worth empirically running the CCA to verify
that the contract is followed, \eg, noise or different topologies does not
cause Copa to mode-switch into Reno's contract.}

The literature has already computed contracts for most CCAs; we cite these in
the ``$\func(\s)$'' column of \autoref{tab:contracts-summary}. This allows us
to focus on the performance impact of contract choice, instead of computing
contracts. Our results hold as long as the assumptions made in the literature
hold (\eg, Copa does not emulate Reno when competing with itself). Even for
CCAs where fluid modeling is hard, their equilibrium behavior is well
understood. For example, BBR~\cite{bbr} encodes fair share in delay when
cwnd-limited~\cite{starvation}, and in ``growth in delivery rate'' when
rate-limited~\cite{bbr-rate-contract}. Contracts can also describe limit-cycle
equilibrium (\eg, sawtooth behavior in Reno/DCTCP) by describing a cycle
instead of a time instance. For instance, DCTCP's contract is
``$\texttt{critical\_cwnd} = 1/(\texttt{avg ECN marking
rate})^2$''~\cite{dctcp}.

For completeness, we show examples of analytical and empirical contract
derivations.


\myparagraph{Analytical contract derivation (Vegas, taken from
\cite{num-srikant})} To update its cwnd, Vegas compares $\texttt{diff} =
\texttt{expected\_throughput} - \texttt{actual\_throughput} = \cwnd/\rtt -
\cwnd/\texttt{RTprop}$ to $\alpha_{\rate}$. It increases cwnd when diff is
larger than $\alpha_{\rate}$, and decreases otherwise. Steady-state occurs when
Vegas has no incentive to change its cwnd, \ie, $\texttt{diff} =
\alpha_{\rate}$. This equation gives us the contract. We substitute $\cwnd =
\rate \cdotp \rtt$, and $\rtt = \texttt{RTprop} + \delay$, and simplify, to get
Vegas's contract: $\rate = \alpha_{\rate} \cdotp \texttt{RTprop}/\delay$. This
has an RTprop bias: flows with larger RTprop get a higher rate. Most
implementations easily remove this by setting $\alpha_{\rate} =
\alpha_{\texttt{pkts}}/\texttt{RTprop}$~\cite{vegas-linux-2.2}. The same can be
done for Reno by removing the RTT term from its contract and implementing the
contract using TFRC~\cite{tfrc}.

\myparagraph{Analytical contract derivation (Astraea, adapted from
~\cite{astraea})} We can compute contracts even for black-box CCAs without
running them. We show this for the RL-based CCA Astraea~\cite{astraea}, which
implicitly learns a contract.
\autoref{fig:astraea} shows its feedback (state) to action mapping. We obtain
this by querying its neural network's action for different feature vectors
(states). The legend shows link capacity in Mbps. For a given capacity and
delay combination (state), action > 1 increases cwnd, and action < 1 decreases
cwnd. Astraea is at a fixed-point when it has no incentive to change cwnd
(action = 0). For each link capacity (or fair share), Astraea maintains a
unique delay, given by the X-coordinates of points with action =
0. We fit a curve ($\rate = a(\delay+b)^c + d$) to these points to get its
contract (\autoref{fig:astraea-fit}). \srini{seems a bit odd to call this
analytic -- seems more like a direct lookup}


\begin{figure}
    \centering
    \includegraphics[]{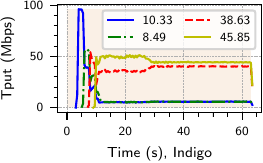} 
    \includegraphics[]{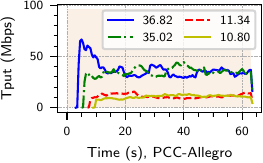} 
    \includegraphics[]{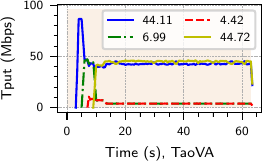} 

    \caption{\label{fig:pantheon-unfair} Indigo~\cite{pantheon},
      Fillp/FillpSheep (TACK~\cite{tack}),
      PCC-Allegro/Vivace/Experimental~\cite{pcc}, and TaoVA (Remy)~\cite{tao,
      remy} are unfair and have no contract. We only show three for brevity. We
    show the time series of throughput of 4 flows. The legend shows the time
  average throughput for the 4 flows.}

\end{figure}

\begin{table*}
    \footnotesize
    \centering
    \begin{tabularx}{\linewidth}{|c|X|c|X|c|X|}
        \hline
        BBR & $({\texttt{avg delay}-46.91})^{-0.89}$ & Vegas & $\texttt{avg delay}^{-1.38}$ & Copa & $\texttt{p50 delay}^{-1.08}$ \\
        \hline
        LEDBAT & $102 - \texttt{p50 delay}$ & Sprout & $134 - \texttt{avg delay}$ & Cubic & $\texttt{loss rate}^{-0.65}$\\
        \hline
    \end{tabularx}

    \caption{\label{tab:pantheon-contracts} Contracts computed for CCAs in
    Pantheon~\cite{pantheon}. We omit constant scaling factors for brevity.}

\end{table*}

\myparagraph{Empirical contract derivation (CCAs in Pantheon)} To demonstrate
that (1) contracts are easy to compute, and (2) fair CCAs have a contract, we
empirically derive contracts for all CCAs in Pantheon~\cite{pantheon} using an
automated procedure. We use Pantheon to run each CCA for
60 seconds on a dumbbell topology with link capacities of 24, 48, and 96 Mbps,
40 ms RTprop, 4 BDP buffer, and vary the flow count from 2 to 8 (except Cubic,
which uses 1 BDP buffer). These yield samples for throughput and three
aggregate statistics: p50 delay (ms), avg delay (ms), and loss rate. We compute
contracts by fitting a curve
to these points (\autoref{fig:pantheon-fit}) and pick the statistic ``\s'' and
fit that minimizes mean squared error.

Of the 17 CCAs in Pantheon, we faced deprecation/dependency issues for QUIC
Cubic and Verus. We run the remainder
15. 2 CCAs (SCReAM and WebRTC) get $\approx 0$ utilization (consistent with the
    Pantheon report~\cite{scream-webrtc}).
    7 CCAs (Indigo, Fillp/FillpSheep, PCC-Allegro/Vivace/Experimental, TaoVA)
  are unfair (\autoref{fig:pantheon-unfair}) and hence, do not have a contract.
  Unfair means different flow rates despite similar statistics/signals,
  indicating the absence of a unique signal-to-rate mapping (contract).
  \autoref{tab:pantheon-contracts} shows the contracts derived for remainder
  6 CCAs.

Our automated procedure is for illustration purposes and is not full-proof.
These contracts are for the range of networks we experiment with. The shifts in
delays (\eg, $46.91$ in BBR and $102$ in LEDBAT) depend on the RTprop or buffer
sizes, \eg, for BBR the shift is equal to RTprop. One would require more
experiments to deduce this empirically. Similarly, the unfair CCAs may be fair
and exhibit contracts on a narrower range of scenarios. For instance, we find
that PCC-Vivace and PCC-Experimental exhibit the contracts ``$\texttt{avg
delay}^{-0.4}$'' and ``$\texttt{p50 delay}^{-0.46}$'' respectively, if we only
consider data from the 96 Mbps link. Finally, the contracts may be compound
functions that take different shapes on different ranges of networks and use
other statistics than the 3 we measured. For instance, \cite{jitu-modelingtcp}
shows Reno's contract under different operating regimes: loss detections are
dominated by timeouts vs duplicate ACKs.


\section{Metrics and tradeoffs}
\label{sec:metrics-tradeoffs}


The choice of contract constrains performance metrics. Conversely, specifying a
desired metric value constrains contract choices, which in turn constrain
the other metrics (\ie, a tradeoff). We begin by defining the metrics in terms
of the contract function \func, so that given \func, the metric value is
obvious. Our metric definitions are unit-less to minimize dependence on the
network parameters (\eg, link capacity, RTprop, \etc). Then, we quantitatively
derive the tradeoffs, \ie, the possible values of a metric if we want another
metric to take some value.

We use running examples of Vegas~\cite{vegas,duality-vegas} ($\x = 1/\s$) and
Swift~\cite{swift} ($\x = 1/\s^2$), where \s is delay. For simplicity, we omit
constant factors that ensure that the units are consistent. \Eg, the unit of
$1$ in Vegas's and Swift's contract is $\text{bytes}$, and
$\text{byte}\cdot\text{seconds}$ respectively, so that the function value has
the unit $\text{bytes}/\text{second}$. \autoref{tab:contracts-summary} shows
where existing CCAs fall in the tradeoff space.


\myparagraph{Metrics} We list the metrics that CCAs typically optimize, and study how contracts affect them.
\vspace{-1em}
\begin{multicols}{2}
\noindent
\setlength{\multicolsep}{0pt}
\begin{packedenumerate}
    \item Link utilization, flow throughput
    \item Amount of congestion (\eg, delay, loss)
    \item Stability, and convergence time/reactivity
    \item Fairness (on general topologies)
    \item Robustness to noise in congestion signals
    \item Generality (range of link rates, flow counts)
\end{packedenumerate}
\end{multicols}
\vspace{-1em}


Of these metrics, fairness notions (see below) subsume Pareto-optimality; that
is, the bottleneck links are fully utilized. They also subsume the tradeoff
between total throughput and fairness (see below).
Stability and convergence time are concerned with transient behavior as opposed
to steady-state equilibrium.
While contracts may affect them, we can often independently optimize them by
choosing how fast a CCA's sending rate moves toward the target rate
(\autoref{sec:canonical}). We are left with the following four metrics:
robustness, fairness, congestion, and generality.

\myparagraph{Generality (\eg, range of link rates supported)} Practical
constraints limit congestion signals to a finite range, $s \in [\Smin, \Smax]$.
Small delays and losses are difficult to measure due to noise, while large
delays and losses harm performance. Since the contract maps congestion signals
to rates, this range confines the CCA's operating range of link rates to
$[\func(\Smax), \func(\Smin)]$, where $\func(\Smin) > \func(\Smax)$ because
$\func$ is decreasing. CCAs must select the constants in $\func$ to suit the
networks they operate on. For instance, if delay spans a $200\times$ range,
\eg, $\delay \in [0.5\ \text{ms}, 100\ \text{ms}]$, Vegas can support only a
$200\times$ range of sending rates (\eg 1 Mbps to 200 Mbps), whereas Swift can
support a $200^2\times$ range (\eg 1 Mbps to 40 Gbps). In general, steeper
contracts enable a broader range of bandwidths for the same domain of the
congestion signal.


\begin{figure}[t]
    \centering
    \begin{minipage}{0.25\textwidth}
        \centering
        \includegraphics[]{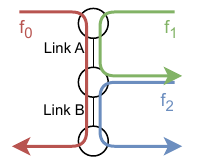}
        \caption{\centering\label{fig:parking_lot} Parking lot topology}
    \end{minipage}\hfill
    \begin{minipage}{0.75\textwidth}
        \footnotesize
        \centering
        \newcolumntype{W}{>{\hsize=.36\hsize\linewidth=\hsize}X}
        \newcolumntype{Z}{>{\hsize=.24\hsize\linewidth=\hsize}X}
        \newcolumntype{Y}{>{\hsize=.2\hsize\linewidth=\hsize}X}
        \begin{tabularx}{\linewidth}{|W|c|Z|c|c|Y|Y|}
            \hline
            \rowcolor{gray!15}
            Fairness notion & $\alpha$ & Global utility
            & $\x[0]$ & $\x[1]=\x[2]$  & Total rate, $\sum_i \x[i]$ & Example CCA\\
            \hline
            \hline
            Max throughput & 0 & $\sum_{i=0}^{k} \x[i]$ & 0 & 1 & 2 &\\
            \hline
            \rowcolor{blue!10}
            Proportional & 1 & $\prod_{i=0}^{k} \x[i]$ & 1/3 & 2/3 & 5/3 & Vegas\\
            \hline
            Min potential delay & 2 & $\sum_{i=0}^{k} -1/\x[i]$ & $\sqrt{2}-1$ & $2-\sqrt{2}$ & $3 - \sqrt{2}$ & Reno\\
            \hline
            \rowcolor{blue!10}
            Max-min & $\infty$ & $\min_i \x[i]$ & 1/2 & 1/2 & 3/2 & Poseidon\\
            \hline
        \end{tabularx}
        \setcaptiontype{table}

        \caption{Throughputs under different fairness notions on the
          parking lot topology (\autoref{fig:parking_lot}). Both links have
          capacity $\bw=1$. $\x[i]$ shows the rate of flow $\f[i]$. From top to
          bottom, fairness (rate equality) improves but total throughput (or
          rate) degrades.
        \label{tab:fairness}}

    \end{minipage}
\end{figure}

\begin{table*}
    \footnotesize
    \centering
    \newcolumntype{Y}{>{\hsize=.20\hsize\linewidth=\hsize}X}
    \newcolumntype{Z}{>{\hsize=.20\hsize\linewidth=\hsize}X}

    \begin{tabularx}{\linewidth}{|Y|c|c|Z|Z|Z|Z|}
        \hline
        \rowcolor{gray!15}
        CCAs & $\func(s) = \sfunc(\s)$ & \agg & Robustness error for \ds noise & Unfairness
        with $k$ hops & Congestion for $n$ flows & Range
        of bandwidths\\\hline
        \rowcolor{gray!15}
        & & & $\max \frac{\sfunc(\s)}{\sfunc(\s+\ds)}$ & $\max
        \frac{\sfunc(\s[k])}{\sfunc(\agg(\s[k], \dots))}$ & $\max
        \frac{\sfunc^{-1}(\bw/N)}{\sfunc^{-1}(\bw)}$ &
        $\frac{\sfunc(\Smin)}{\sfunc(\Smax)}$\\
        \hline
        \rowcolor{gray!15}
        & & & Lower is better & Lower is better & Lower is better & Higher is
        better\\
        \hline
        \rowcolor{gray!15}
        & & & Want gradual \sfunc & Want gradual \sfunc & Want steeper \sfunc &
        Want steeper \sfunc \\
        \hline
        \hline
        \rowcolor{blue!10}
        FAST, Copa & $1/\s$~\cite{fasttcp, num-low, starvation} & $\sum$ & $1 +
        \ds/\Smin$ & $k$ & $n$ & $\frac{\Smax}{\Smin}$\\
        \hline
        Swift, DCTCP & $1/\s^2$~\cite{swift, dctcp, dctcp-analysis} & $\sum$ & $(1 +
        \ds/\Smin)^2$ & $k^2$ & $\sqrt{n}$ &
        $\left(\frac{\Smax}{\Smin}\right)^2$\\\hline
        \rowcolor{blue!10}
        Reno & $1/\sqrt{\s}$~\cite{reno,mathis-model, num-low} & $\sum$ & $\sqrt{1 +
        \ds/\Smin}$ & $\sqrt{k}$ & $n^2$ & $\sqrt{\frac{\Smax}{\Smin}}$\\\hline
        Poseidon & $e^{-\s}$~\cite{poseidon} & $\max$ & $e^{\ds}$ & $1$ & $\log{n}$ &
        $e^{\Smax - \Smin}$\\\hline\hline
        \rowcolor{green!10}
        $\alpha$-fair (\autoref{sec:metrics-tradeoffs})
        & $1/\sqrt[\alpha]{\s}$~(\autoref{eq:utility}) & $\sum$ & $\sqrt[\alpha]{1 +
        \ds/\Smin}$ & $\sqrt[\alpha]{k}$ & $n^{\alpha}$ &
        $\sqrt[\alpha]{\frac{\Smax}{\Smin}}$\\\hline
        Exponential & $e^{-s/\Sig[0]}$~\cite{starvation} & $\sum$ & $e^{\ds/\Sig[0]}$ & $e^{k \cdotp
        \Smax/\Sig[0]}$ & $\log{n}$ & $e^{(\Smax - \Smin)/\Sig[0]}$\\\hline\hline
        \rowcolor{violet!10}
        ICC & $\log(\Sig[0]/\s)$~\cite{icc} & $\sum$ & $\infty$ & $\infty$ & $\left(\frac{\Sig[0]}{\Smin}\right)^{\frac{n-1}{n}}$ &
        $\frac{\log{\Sig[0]} - \log{\Smin}}{\log{\Sig[0]} - \log{\Smax}}$\\\hline
        Astraea & $\bw[0](1-\s/\Sig[0])$ (\autoref{sec:contracts}) & $\sum$ & $\infty$ & $\infty$ &
        $\frac{n\bw[0]-\Cmax}{n(\bw[0]-\Cmax)}$ &
        $\frac{\Sig[0]-\Smin}{\Sig[0]-\Smax}$\\\hline
        \rowcolor{violet!10}
        AIMD on delay & $1/\sqrt{\s}$ (\autoref{sec:pitfalls}) & (\autoref{sec:pitfalls}) & $\sqrt{1 +
        \ds/\Smin}$ & $\infty$ & $n^2$ & $\sqrt{\frac{\Smax}{\Smin}}$\\\hline
    \end{tabularx}

    \caption{\label{tab:contracts-summary} The ``periodic table'' of CCAs and the tradeoff space.
      We organize into 3 sections: good contracts (\hlblue{blue}, top), corner
      points (\hlgreen{green}, middle), and bad contracts (\hlviolet{pink},
      bottom). For Reno, \s is loss rate. For DCTCP, \s is ECN marking rate.
      For Poseidon \s is max per-hop delay. For AIMD on delay, \s is bytes
      between high delay. For all other CCAs \s is delay. We assume loss and
      ECN marking rates are small enough to approximate \agg as $\sum$.
      For ICC and Astraea, unfairness and robustness error are worst when $\s
      \approx \Sig[0]$.
    }

\end{table*}


\myparagraph{Robustness to noise} We study the impact of ``\ds'' error in the
statistic ``\s'' on the CCA's rate. We quantify the impact by looking at the
(worst-case) ratio of rates without and with the noise: $\text{Error factor} =
\max_{\s} \frac{\func(\s)}{\func(\s+\ds)}
~\refstepcounter{equation}(\theequation)\label{eq:error-factor-def}$. A higher
error factor means that the CCA is more sensitive to noise and is less robust.
From \autoref{eq:error-factor-def}, the error factors for Vegas ``$\x = 1/\s$''
and Swift ``$\x = 1/\s^2$'' are ``$1 + \ds/\Smin$'', and ``$\left(1 +
\ds/\Smin\right)^2$'' respectively. If \ds error perturbs Vegas by $2 \times$,
then the same error perturbs Swift by $4\times$. Steeper contracts yield higher
(worse) error factors (\autoref{fig:contracts:robustness}).

Note that the amount of noise we want to tolerate may depend on the statistic.
\Eg, the delay may have tens of milliseconds of noise~\cite{abc}, while the
loss rate may have a noise of a few percent~\cite{tcp-random-loss}.

\myparagraph{Fairness} For general network topologies, there exist multiple
notions of fairness (\autoref{tab:fairness}). These are typically parameterized
using $\alpha$-fair utility functions~\cite{alpha-fairness} given by:
$\utility_\alpha(\x) = \frac{\x^{1-\alpha}}{1-\alpha}$, where $\utility$
represents the utility a flow derives from a rate of $\x$, and $\alpha > 0$.
The rate allocation maximizes the sum of utilities across all flows
$\left(\sum_i \utility_{\alpha}(\x[i])\right)$ (global utility).
  Larger values of $\alpha$ indicate greater fairness at the cost of total
  throughput~\cite{throughput-vs-fairness}. Typically, $\alpha \geq 1$ is
  desired.

Not all contracts correspond to an $\alpha$-fair utility. As a proxy, to
compare the fairness of contracts, we study their behavior in the parking lot
topology (\autoref{fig:parking_lot}). This topology exposes differences between
fairness notions (\autoref{tab:fairness}), and is common in data centers (\eg,
when inter- and intra-rack flows compete). In \autoref{fig:parking_lot}, each
link has a capacity of \bw bytes/second and flows experience congestion signals
from multiple congested links. The \emph{long} flow ($\f[0]$) observes signals
from two hops, while the \emph{short} flows ($\f[1]$ and $\f[2]$) observe
signals from only one hop. More generally, for $k$ hops, we consider flows
$\langle \f[0], \f[1], \ldots, \f[k] \rangle$, where $\f[0]$ observes signals
from all $k$ hops, and other flows see signals from a single hop.

We quantify unfairness as the (worst-case) throughput ratio of $\f[k]$ to
$\f[0]$. We derive unfairness using three steady-state equations. First, at
each hop, the sum of the incoming rates is equal to the link capacity: $\forall
i \geq 1, \, \x[0] + \x[i] =
\bw~\refstepcounter{equation}(\theequation)\label{eq:fairness:balance}$.
Second, each flow's rate (\x[i]) and statistic (\s[i]) follow the contract
$\forall i \geq 0, \, \x[i] =
\func(\s[i])~\refstepcounter{equation}(\theequation)\label{eq:fairness:contract}$,
Third, \f[0] sees an accumulation of signals from all the hops: $s_0 =
\agg(s_1, s_2, \ldots,
s_k)~\refstepcounter{equation}(\theequation)\label{eq:fairness:agg}$. Here,
\agg describes the statistic $\f[0]$ sees when other flows see $s_1, s_2,
\ldots, s_k$. For delay, $s_0 = \sum_{i=1}^{k} s_i$, as delays add up over
hops. For loss rate and ECN marking rate, $s_0 = 1 - \prod_{i=1}^k (1-s_i)$, as
survival probability gets multiplied over the hops. Note, when the absolute
value of loss rates (or ECN marking rates) is small (\eg, $10^{-2}$), then \agg
can be approximated as a sum. For max per-hop delay (\eg,
Poseidon~\cite{poseidon}), $s_0 = \max(s_1, s_2, \ldots, s_k)$.




From \autoref{eq:fairness:balance}, we get $\x[1] = \x[2] = \dots \x[k]$.
Given, \func is invertible, we get $\s[1] = \s[2] = \dots = \s[k]$. Then, the
worst-case throughput ratio is given by: $\max_{\bw} \frac{\x[k]}{\x[0]} =
\frac{\func(\s[k])}{\func(\s[0])} =
\max_{\bw}\frac{\func(\s[k])}{\func(\agg(\s[k], \s[k], \dots, k \text{
times}))}~\refstepcounter{equation}(\theequation)\label{eq:fairness-def}$.
Here, $\s[k]$ solves $\func(\s[k]) + \func(\agg(\s[k], \s[k], \dots, k \text{
times})) = \bw$ (from \autoref{eq:fairness:balance}).

Substituting the contracts for Vegas and Swift, and \agg as $\sum$, we get the
throughput ratios of $k$ and $k^2$ respectively. Steeper contracts imply worse
fairness.


\myparagraph{Congestion growth} We study how congestion grows with (decreasing)
fair share or (increasing) flow count. Consider a dumbbell topology with
capacity \bw and $n$ flows. For each flow to get its fair share of $\x[i] =
\bw/n$, the statistic needs to satisfy: $\forall i$, $\x[i] = \bw/n =
\func(\s[i])$. This implies that $\func^{-1}(\bw/n) = \s[n] = \s[i]$. We define
worst-case growth (\ie, (\s for n flows)/(\s for 1 flow)) as: $\grow(n) =
\max_{\bw}
\frac{\func^{-1}(\bw/n)}{\func^{-1}(\bw)}~\refstepcounter{equation}(\theequation)\label{eq:growth-def}$.
For Vegas and Swift, $\grow(n)$ is $n$ and $\sqrt{n}$ respectively. Steeper
contracts imply slower growth.





\myparagraph{Tradeoff summary} Gradual contracts give better robustness and
fairness, while steeper contracts give better congestion and generality. In
\autoref{app:tradeoffs}, we mathematically show these tradeoffs by considering
each pair of contending metrics. \Eg, if we want error factor $\leq \ef$ for
\ds noise, then $\grow(n)$ is
$\Omega\left(\ds\frac{\log(n)}{\log(\ef)}\right)$. These bounds are tight, \ie,
we show CCAs that meet these bounds (see below).

The tradeoffs involving fairness depend on \agg. We consider $\agg \in \{\sum,
\max, \min\}$. The tradeoff exists for $\agg=\sum$, \eg, delay, loss rate, and
ECN marking rate (assuming loss and ECN marking rates are small enough). There
is no tradeoff for $\max$ and $\min$, \eg, max per-hop delay, as this allows
independently ensuring max-min fairness. So, for CCAs like
Poseidon~\cite{poseidon}, the only tradeoff is between robustness versus
congestion and generality.



There are exactly two corner points in this tradeoff space. If we fix desired
robustness, \eg, we want to tolerate \ds noise, then the Exponential
CCA~\cite{starvation} gives the best generality and congestion growth, but it
has poor fairness (\autoref{tab:contracts-summary}). If we fix desired
fairness, \eg, proportional fairness ($\alpha=1$), then Vegas gives the best
generality and congestion growth (for $\agg = \sum$). In general, if we want
$\alpha$-fairness, then the contract ``$\func(\s) =
\frac{1}{\sqrt[\alpha]{\s}}$'' gives the best generality and congestion growth
(for $\agg = \sum$). We mathematically show that these are corner points in
\autoref{app:tradeoffs}.


\subsection{Guidance on picking contract (D1)}

\myparagraph{(D1.1) Input/output} Unless one chooses a workaround to the
tradeoffs (\autoref{sec:workarounds}), the choice of contract input is one of
delay, loss rate, or ECN marking rate. This choice depends on buffer sizes,
availability of ECN, and constraints on dynamics (\autoref{sec:canonical}).
Similarly, outside of the workaround of coordinating fraction of link use
(\autoref{sec:workarounds}), the output of the contract is rate.


All the workarounds in \autoref{sec:workarounds} either require in-network
support (\eg, max per-hop-delay signal) or fundamental congestion control
research. For instance, \cite{starvation} shows that to ensure robustness and
full generality (arbitrarily large link rates), CCAs must deliberately vary
rate and create delay variations. This goes against our desire to have a stable
application-level rate. In fact, RL-based CCAs~\cite{astraea} and
Remy~\cite{remy} update cwnd based on moving averages of historical signals.
CCAs that are deterministic responses to such histories are incapable of
creating deliberate rate/delay variations.

%
%


\myparagraph{(D1.2) Shape} For the above input/output choices, the tradeoffs
exist. One should decide which of the two corner points in the tradeoff space
is preferred.
The tradeoff point fixes the contract shape (\eg, $\rate = 1/\s$) and the
asymptotics in the tradeoffs.
Next, we want to tune the contract for deployment, \ie, picking parameters that
control shift or scaling (\eg, $a, b, c$ in $\rate = a / (\s - b) + c$). The
scale (\eg, $a$) affects the constant factors in the tradeoffs, and
shifts/clamps (\eg, $b, c$) should be avoided.

\myparagraph{(D1.3) Shift/clamps} Shifting the contract right (positive $b$)
changes the fairness properties (\autoref{sec:pitfalls}). Shifting left
(negative $b$) restricts the range of link capacities by introducing a
Y-intercept. Shifting down (negative $c$) introduces an X-intercept which
severely degrades robustness and fairness (\autoref{sec:pitfalls}). Shifting up
(positive $c$) only makes sense if there is a known lower bound on the fair
share of a flow (\eg, fair share $\geq 512$ Kbps). Clamping has similar
consequences.



\myparagraph{(D1.3) Scale} We discuss the scaling choice for the two corner
contracts. We use \bw[0] and \Sig[0] respectively to represent scaling of \rate
and \s, \eg, $\rate = \bw[0] e ^{-\s/\Sig[0]}$. One can set these parameters to
meet a desired value for one of the four metrics. The tradeoffs decide the
other metrics.

The general form of the exponential contract is $\rate = \bw[0] e
^{-\s/\Sig[0]}$. \bw[0] is the maximum rate at which the CCA can ever send. If
the link capacity is higher, the CCA would under-utilize the link. \Sig[0]
controls the robustness error (throughput ratio) for \ds error in \s $= \Cmax
e^{-\s/\Sig[0]}/\Cmax e^{-(\s+\ds)/\Sig[0]} = e^{\ds/\Sig[0]}$. Larger \Sig[0]
gives better robustness but worse congestion. These parameters also decide the
range of delays (\Smin, \Smax) produced by the CCA for a deployment's range of
fair shares (\Cmin, \Cmax).

The general form of the $\alpha$-fair contract is $\rate = \bw[0]
{\Sig[0]}^{\alpha} / \s^{\alpha}$, or $\rate = \bw[0]\Sig[0]/\s$ for
$\alpha=1$. The contract passes through $\langle\Sig[0], \bw[0] \rangle$:
\Sig[0] is the delay maintained when the fair share is \bw[0]. The deployment's
range of fair shares ($\Cmin, \Cmax$) implicitly decides the range of
congestion ($\Smin, \Smax$), and robustness to noise. For small capacities
($\Cmin \to 0$), \Smax is $\infty$. For large capacities ($\Cmax \to \infty$),
\Smin is 0 and the error factor is $\infty$ (as $\ds/\Smin \to \infty$). When
\Cmax is known (\eg, a data center), and one desires to maintain a minimum delay
of \Smin (to ensure utilization despite variations~\cite{astraea}, or bound
error factor by ``$1 + \ds/\Smin$''), an example contract choice is $\rate =
\Cmax\Smin/\delay$.


The exponential contract ensures a finite error factor at the cost of not
supporting arbitrarily large link capacities. The $\alpha$-fair contract
supports arbitrarily large capacities but cannot bound the error factor. This
is the same as the fundamental tradeoff in~\cite{starvation}.





\begin{figure*}
    \centering
    \includegraphics[]{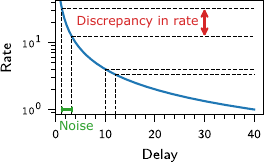}
    \includegraphics[]{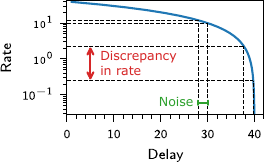}
    \includegraphics[]{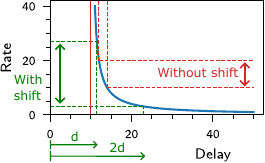}

    \caption{\label{fig:error-shift} \textbf{(Left)} Vegas's contract ($\rate =
      1/\delay$). The same noise creates larger discrepancies with increasing
      link capacities. \textbf{(Middle)} Astraea's contract ($\rate = \Sig[0] -
      \delay$) has an X-intercept. Discrepancies increase as delay approaches
      the intercept. \textbf{(Right)}
    Shifting the contract changes the fixed-point delays and rates. We omit
  units (\eg, Mbps or ms) here as the shape/shift/intercept matter not the
scale.}

\end{figure*}

\section{Learnings from contracts}
\label{sec:pitfalls}

\anup{Most CCAs overlook robustness and multi-bottleneck fairness. That's where
most of the learnings lie.}

\myparagraph{Avoid extreme shape and intercepts} Astraea (linear contract) and
ICC (logarithmic contract) exhibit poor robustness and fairness due to extreme
shapes and X-intercepts. The issue is that all the low rates (\eg, 0.1 to 1
Mbps) map to delays near the X-intercept (\eg, 40 ms)
(\autoref{fig:error-shift} middle). Consequently, small delay jitter creates
large discrepancies in inferring fair shares. Similarly, on parking lot
(\autoref{fig:parking_lot}), if the hop delays are close to the intercept
delay, the long flow (\f[0]) observes their sum, which exceeds the intercept
delay. Attempting to reduce this ``excessive delay'', \f[0] reduces its rate to
zero and starves.

\begin{figure}
  \centering
  \begin{minipage}{0.49\textwidth}
    \centering
    \includegraphics[]{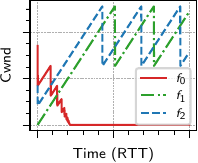} 
    \includegraphics[]{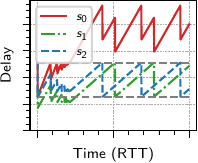}
    \caption{\centering\label{fig:starvation-aimd-delay} AIMD on delay starves the long
    flow (\f[0]) on parking lot topology.}
  \end{minipage}
  \begin{minipage}{0.49\textwidth}
    \centering
    \includegraphics[]{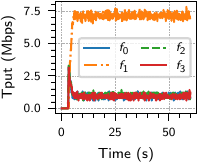} 
    \includegraphics[]{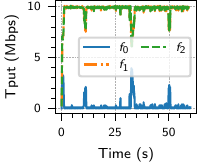}
    \caption{\centering\label{fig:extreme-contract-bad} Astraea (left) is brittle under
    ambient emulation noise and ICC (right) starves on parking lot.}
  \end{minipage}
\end{figure}

\autoref{fig:extreme-contract-bad} demonstrates these issues empirically.
Ambient emulation noise causes Astraea to starve flows when the fair share is
low (4 flows on a dumbbell topology with 10 Mbps capacity, 30 ms RTprop, and
infinite buffer). ICC starves the long flow (\f[0]) on a parking lot topology
with 3 flows (2 hops) and the same network parameters. Importantly, contract
analysis reveals these performance issues without running the CCAs or
understanding their internal working.

Note, unlike \cite{starvation}, the starvation here is not fundamental and is
easily avoided by removing the X-intercept (\eg, \rate = 1/\delay). This
increases the delay at low fair shares, however, this is unavoidable as
transmission delays anyway grow as $1/\bw$ with decreasing capacity. In
contrast, the issue in \cite{starvation} is fundamental and is caused by the
asymptote on the Y-axis, where all the high fair shares (\eg, > 100 Mbps) map
to near-zero delays (\autoref{fig:error-shift} left). This asymptote is hard to
remove while supporting arbitrarily large rates with a monotonically decreasing
contract.

\myparagraph{Avoid shifting contracts} CCAs like Swift~\cite{swift} define
target delay as: $\texttt{target\_delay} = b + 1/\rate$. In such formulations,
it may seem convenient to shift the contract (set a positive $b$) to maintain a
minimum delay for high utilization. However, shifting changes the steady-state
fixed-point causing undesired unfairness (\autoref{fig:error-shift} right). For
instance, in the parking lot topology with 2 hops, the short and long flows see
delays of $d$ and $2d$ respectively. These delays are such that the
corresponding contract rates add up to the link capacity. Shifting the contract
rightward changes these delays in a way that increases discrepancies in the
rates. We show this empirically for BBR in \autoref{sec:empirical}. To avoid
this issue, one should tune the scaling (\autoref{sec:metrics-tradeoffs}) to
maintain a minimum delay instead.



With active queue management (AQM), \eg, RED-based ECN marking~\cite{red, ecn},
we can increase queue buildup (delay) independent of the contract. We can
increase the RED parameters $K_{\min}, K_{\max}$ to increase delay without
changing the steady-state ECN marking probabilities (and rate allocations).

\myparagraph{Avoid fixed thresholds (\eg, delay targets) at end-hosts} We
illustrate how AIMD on delay (\ie, MD when delay crosses a fixed threshold and
AI otherwise) causes starvation. AIMD on delay appears in 1RMA~\cite{1rma} and
the CCAs (SMaRTT~\cite{smartt} and STrack~\cite{strack}) proposed for
standardization in the Ultra Ethernet Consortium~\cite{uec}. It was also
proposed by \cite{starvation} to work around their impossibility result.


\anup{Swift at low rates also does AIMD on delay.}

On a dumbbell topology, AIMD on delay works fine. It induces a Reno-like
contract: ``$\x = 1/\sqrt{\s}$'', where \s is the high-delay probability
(inverse of bytes between high-delay events) instead of loss probability.
However, it creates starvation with multiple bottlenecks. All hosts attempt to
maintain delay around the same (fixed) target value, but when flows observe
delays from different combinations of hops, they cannot simultaneously maintain
the same delay. \autoref{fig:starvation-aimd-delay} shows this on a 2-hop
parking lot. The short flows (\f[1] and \f[2]) oscillate delay between the
delay threshold and half the threshold (shown in \textcolor{TextGray}{gray}).
The long flow (\f[0]) observes their sum, which always exceeds the threshold.
Consequently, \f[0] never increments cwnd and starves.

We can resolve this by either (1) not using a fixed threshold, \eg, Swift
scales the (target delay) threshold with rate, which transforms the contract to
use delay instead of high-delay probability, allowing different flows to
maintain different delays, or (2) moving the threshold to links instead of
end-hosts, \eg, packet drops (Reno) and ECN marks (DCTCP) occur when delay
crosses a threshold at the links.

This issue may arise for any CCA that uses fixed end-to-end thresholds to
decide when to increase or decrease their rate, \eg, BBRv3 uses
$\texttt{BBRLossThresh}=2\%$~\cite{bbr-ietf}.

\myparagraph{Minor changes in CCA change the contract and consequently
steady-state performance} MPRDMA~\cite{mprdma}, DCTCP~\cite{dctcp}, and
Reno~\cite{reno} all perform AIMD on binary feedback. In
fact, \cite{htsim} uses MPRDMA as an approximation of DCTCP. However, the minor
differences in their design result in different contracts: Reno: ``$\x =
1/\sqrt{\s}$'', DCTCP: ``$\x = 1/\s^2$'', MPRDMA: ``$\x = 1/\s$''.

Similarly, we found an algebraic mistake in the Linux implementation of TCP
Vegas that changes its contract from ``$\x = 1/\delay$'' to ``$\x =
1/\delay^2$'' when RTprop is small. We have confirmed this with the
maintainers. This bug has existed for
17 years due to a refactoring commit~\cite{vegas-bug-commit}. Such bugs may be
   caught immediately if CCA implementations explicitly delineate contracts.

\section{Canonical CCA dynamics to follow a contract (D2)}
\label{sec:canonical}

\begin{SCfigure}
  \centering
  \includegraphics[]{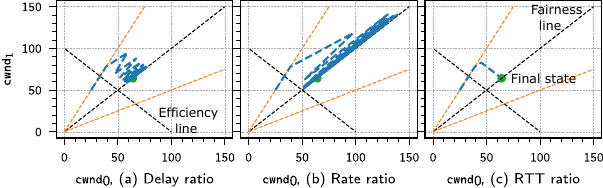}

  \caption{
  MIMD using RTT ratio is more stable than rate or delay ratio. Dynamics stay
above the efficiency line showing no throughput loss, contrary to PowerTCP's
claim on voltage-based CCAs~\cite{powertcp}. \label{fig:dynamics}}

\end{SCfigure}

We can implement a rate or cwnd-based contract using either rate or cwnd. We
discuss cwnd-based CCA to follow a rate-based contract. Other combinations are
similar. Existing CCAs correspond to various ways of implementing cwnd updates
to follow a contract. While the contract fixes steady-state performance, the
updates determine dynamics (stability and convergence time). Vegas uses AIAD,
Swift uses AIMD, Posiedon uses MIMD (using rate ratio), Copa uses AIAD with
increasing gains when cwnd updates occur in the same direction, and FAST uses
(a different kind of) MIMD (using RTT ratio).

We argue that TCP FAST's method is the best way to implement a contract.
AIMD/AIAD are sub-optimal in convergence time. Copa
uses convex---instead of concave (\eg, ETC~\cite{etc-atc})---changes to cwnd,
which creates overshoots and instability. Ideally, we want to dampen changes to
cwnd when it is already close to convergence~\cite{etc-atc}. Poseidon's MIMD
creates instability depending on network parameters (see below). FAST's update
is stable and converges exponentially fast to both efficiency and fairness.
We illustrate (\autoref{fig:dynamics}) the issue with Poseidon's MIMD and how
FAST's MIMD fixes the issue in the context of delay, and later discuss other
congestion signals.


\myparagraph{MIMD using rate or delay ratio (C.f. Poseidon)} We can interpret a
contract (\eg, $\rate = 1/\delay^2$), in two ways: (1) \emph{target rate} given
\emph{current delay} ($\texttt{target\_rate} = 1/\texttt{current\_delay}^2$),
or (2) ``\emph{target delay} as a function of \emph{current rate}''
($\texttt{target\_delay} = 1/\sqrt{\texttt{current\_rate}}$). Where,
$\texttt{current\_rate} = \cwnd/\rtt$, and $\texttt{current\_delay} = \rtt -
\min \rtt$. These yield two natural MIMD updates to implement a contract:
\begin{align}
  & \text{(1) } \tcwnd \gets \cwnd \frac{\texttt{target\_rate}}{\texttt{current\_rate}} \quad
  \text{(2) } \tcwnd \gets \cwnd \frac{\texttt{target\_delay}}{\texttt{current\_delay}}
\end{align}
Where cwnd moves towards the target with optional clamps bounding the magnitude of
change: ``$\texttt{next}\_\cwnd \gets (1-\cwndupdateavg) \cdotp \cwnd +
\cwndupdateavg \cdotp \tcwnd$'' and ``$\cwnd \gets
\texttt{clamp}(\texttt{next}\_\cwnd, \beta_{\min}\cwnd, \beta_{\max}\cwnd)$''.
Outside of ``\rate = 1/\delay'', rate and delay ratios are different yielding
different dynamics. For each \cwndupdateavg, clamp choice, both updates are
only stable for specific network parameters (\eg, capacity, flow count).


\myparagraph{MIMD using RTT ratio (C.f. FAST)} To ensure stability, we want to
consider how rate and delay affect or are affected by changes in cwnd. The
following update achieves stability without requiring any averaging
(\cwndupdateavg), clamps, or assumptions on network parameters:
\begin{align}
      & \cwnd \gets \cwnd \cdot \frac{\texttt{target\_RTT}}{\texttt{current\_RTT}} = \cwnd \cdot \frac{\min \rtt + \texttt{target\_delay}}{\min \rtt + \texttt{current\_delay}}
      \label{eq:stable-mimd}
\end{align}
Intuitively, cwnds map to ``packets in queue + packets in pipe''. We only want
to move the ``packets in queue'' part from the current delay to the target
delay. \autoref{eq:stable-mimd} isolates the term responsible for queueing
delay and only scales that term using the delay ratio. We can also interpret
this as: $\cwnd \gets \rate * \textcolor{blue}{\texttt{target}\_\texttt{RTT}} =
\cwnd/\rtt * \textcolor{blue}{(\min \rtt + \texttt{target\_delay})}$.


\smallskip\noindent\textbf{For other signals,} \eg, ECN or loss, the best cwnd
update may differ. The current and target RTT in the update depends on the
relation between RTT and aggregate statistic. For instance, for RED-based ECN,
given \texttt{target\_ECN\_rate}, the \texttt{target\_RTT} is ``$ \min \rtt +
(1/\bw) \cdotp (K_{min} + (K_{max} - K_{min}) * \texttt{target\_ECN\_rate})$''.
We derive this by inverting the mapping from queue size (and hence queueing
delay) to ECN marking probability. This works directly when ground truth link
capacity ``\bw'' is known (\eg, a data center), otherwise either the capacity
needs to be estimated or one needs to use delay or rate ratio with a value of
\cwndupdateavg tuned for the range of network parameters one wants to support.

TFRC~\cite{tfrc} explores cwnd update choices for loss-based contracts. The
challenge is that while a constant cwnd creates constant delay and ECN rate, a
constant cwnd may not create a constant loss rate (at least when loss rate is
less than one loss per window). Thus, we may need cwnd variations to maintain
a persistent loss rate even when target and current loss rates are the same.

\srini{I guess this is arguing that TFRC was designed in the context of most
flows using RENO to create the loss rate. A pure TFRC network wouldn't work
without introducing cwnd variations into the protocol. Am I interpreting this
correctly? If so, do the cwnd updates considered in that paper really make
sense?} \anup{I assumed that TFRC does the correct thing. I am trying to
communicate that simple MIMD may not work for loss, give intuition why, and
refer to TFRC for solutions. Yes, to my recollection, TFRC did not evaluate
cases where only TFRC ran or when there are no non-congestive losses. So it is
hard to say if it induces losses correctly.} \srini{TBH, I don't recall their
exact approach but do recall thinking it made no sense in isolation} \anup{I
guess i just say that TFRC explore this and not whether its correct and explain
that loss is more complicated than delay. You think this para is not clear
enough?} \srini{I guess it delivers that -- I would consider dropping the TFRC
citation -- since it makes it sound like there is a solution there, but I am
not convinced there is. }

\myparagraph{Other design considerations} A complete the CCA needs other
decisions including: (O1) how to aggregate multiple statistic samples, (O2) how
long to measure the samples, (O3) time between cwnd updates, and (O4) how to
compute any other estimates (\eg, bandwidth or RTprop estimate). For instance,
Copa computes standing RTT by taking the ``minimum'' (O1) over RTT samples in
the last ``half srtt'' (standard smoothed RTT) (O2). It updates cwnd ``every
ACK'' (O3) and uses minimum RTT over last
10 seconds to estimate RTprop (O4). Alternatively, one can estimate RTprop
   using BBR's RTT probes.

\section{Empirically validating the metrics/tradeoffs}
\label{sec:empirical}


\begin{figure*}
    \centering
    \includegraphics[width=0.32\textwidth]{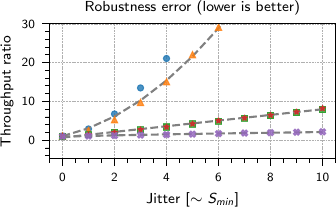}\hfill
    \includegraphics[width=0.32\textwidth]{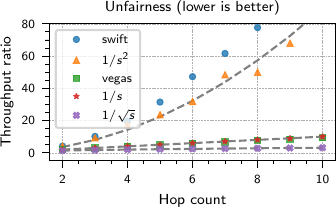}\hfill
    \includegraphics[width=0.32\textwidth]{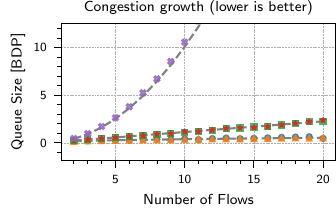}\hfill

    \caption{\label{fig:validation} Contracts-based performance estimates match
      packet-level simulation. The markers show empirical data and
      \textcolor{TextGray}{gray} lines show performance estimated by contracts.
      \autoref{fig:app:validation} in \autoref{app:validation} shows log-log
    scale. Note: markers may be hard to see because they overlap.}

\end{figure*}

We empirically validate the trends in metrics and tradeoffs predicted by
contracts. For visual clarity in plots, we only show a handful of
contracts/CCAs. This section complements the empirical results in
\autoref{sec:contracts}, \autoref{sec:metrics-tradeoffs}, and
\autoref{sec:pitfalls}; where we already showed performance issues and
contracts for a large set of CCAs including Sprout, PCC, Indigo, ICC, Astraea,
AIMD on delay (1RMA, SMaRTT, STrack, \etc).

\myparagraph{Methodology} We use packet-level simulation and emulation.
Simulations allow us to control noise and isolate one source of tradeoff at a
time where emulation always has ambient noise, \eg, due to OS scheduling
jitter. We use htsim~\cite{htsim} for simulation (previously used in
MPTCP~\cite{mptcp}, NDP~\cite{ndp}, EQDS~\cite{eqds}), and
mahimahi~\cite{mahimahi}, Pantheon~\cite{pantheon}, and mininet~\cite{mininet}
for emulation.

In simulation only, for two reasons, we give oracular knowledge of RTprop to
all CCAs. First, CCAs like Swift target data center deployments where RTprop
may be known. For consistency, we provide RTprop to all CCAs. Second, CCAs like
Vegas do not explicitly drain queues resulting in misestimating RTprop and poor
performance. We remove this source of poor performance as this is not
fundamental unlike the tradeoffs imposed by contracts. For instance, Copa and
BBR explicitly drain queues to estimate RTprop accurately (at least in the
absence of noise).

Note that we are validating negative results (\ie, tradeoffs). Simplifications
only make the validation stronger. If tradeoffs exist with oracular knowledge
of RTprop, then performance is only worse without it, \eg, under-estimation
causes under-utilization, and over-estimation increases congestion.

\myparagraph{Simulation CCAs} We implement and test Swift~\cite{swift} and
Vegas~\cite{vegas}. For reference, we also show 3 canonical
(\autoref{sec:canonical}) contract implementations: $1/\sqrt{\s}$, $1/\s$, and
$1/\s^2$, where $\s=\delay=\rtt - \texttt{RTprop}$. These avoid RTT-bias unlike
vanilla Vegas/Swift, and use MIMD instead of AIAD/AIMD. In the canonical CCAs,
we update cwnd every 2 RTTs and aggregate delay as the minimum over delay
samples since the last cwnd update. To isolate the impact of contract shape,
we also tune the scale parameters symmetrically. All CCAs have the same
$\Smin=1.2$ $\mu s$ $= 0.1$ RTprop and $\Cmax=100$ Gbps = link capacity. \Smax
is then decided by \Cmin and the contract shape.

\myparagraph{Simulation scenarios} We defined the metrics in
\autoref{sec:metrics-tradeoffs} to be unitless. The tradeoffs exist for all
choices of network parameters (link capacity, RTprop, \etc) and their absolute
values have little consequence. We set link capacity $= 100$ Gbps, RTprop $= 12$
$\mu s$, packet size $= 4$ KB. Since we use delay-based CCAs, we set the buffer
size to be infinite to remove any effects of packet losses. To measure
robustness error, we use a dumbbell topology with 2 flows where one of
them witnesses noise. We inject controlled error by
adding a hop that persistently delays packets by ``\ds'' $\mu s$, and vary \ds.
\srini{is this one-sided error -- i.e. RTProp is always smaller than actual
transmission delay?}\anup{Not sure I follow what you mean and what is the
purpose of this question. how does this change the text?} We do not include this
in the RTprop provided to the CCAs. We inject noise this way to show trends. In
emulation, we show the impact of realistic noise. To measure unfairness and
congestion growth, we instantiate parking lot (with varying hops) and dumbbell
(with varying flow count) topologies respectively. For the CCAs we test,
generality is just the inverse of congestion growth
(\autoref{tab:contracts-summary}), so we do not show generality.


\myparagraph{Simulation results} Empirical performance matches that estimated
by contracts (\autoref{fig:validation}). Swift's RTT-bias causes slightly worse
fairness and robustness than the equivalent ``$1/\s^2$'' canonical CCA which
removes the bias. Vanilla Vegas has RTprop-bias (instead of RTT-bias). Since
the RTprop is the same for all flows, the steady-state performance of Vegas is
the same as the canonical ``$1/\s$'' CCA.

\begin{figure*}
    \centering
    \includegraphics[]{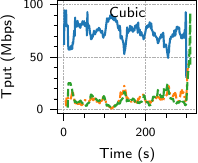}
    \includegraphics[]{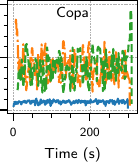}
    \includegraphics[]{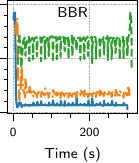}\hfill
    \includegraphics[]{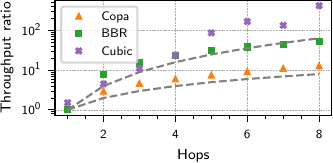}

    \caption{\label{fig:kernel-validation} \textbf{Robustness error (Left 3).}
      Cubic starves the orange/green flows that do not witness jitter, while
      Copa starves the blue flow that witnesses jitter. BBR starves the blue
      flow with the smallest (10 ms) RTprop. \textbf{Unfairness (Right).} Grey
    lines show $y=x$ and $y=x^2$. Copa matches the unfairness predicted by its
  contract. BBR and Cubic are worse due to shift and RTT-bias.}

\end{figure*}

\myparagraph{Emulation CCAs, scenarios, and results} We run Cubic~\cite{cubic},
BBR~\cite{bbr} (Linux kernel v5.15.0) and Copa~\cite{copa,genericCC}. The
empirical contract derivations (\autoref{sec:contracts}) already showed
generality and congestion growth, \eg, increase in delay or loss rate with
decreasing fair share (increasing flow count). In
\autoref{fig:kernel-validation}, we show robustness and fairness. We run flows
for 5 mins on dumbbell and parking lot topologies with capacity $= 100$ Mbps,
buffer $= 1$ BDP, and describe RTprop and flow count below.

Note that introducing noise in signal may not create an equivalent amount of
error in the statistic used by the CCA. Hence, we do not see a persistent trend
in robustness error with varying noise. To validate that robustness is an
issue, we show that the CCAs incur large throughput ratios (starvation) with
small delay jitter. We inject jitter in two ways: (J1) slightly different
RTprops (3 flows with RTprop of 10, 20, and 30 ms), and (J2) ACK-aggregation (3
flows with RTprop of 32 ms but 1 flow additionally witnesses
32 ms of ACK aggregation). We emulate ACK-aggregation in the same way as
   Pantheon~\cite{pantheon}. In \autoref{fig:kernel-validation} (Left 3), Cubic
   and Copa show starvation with J2 and BBR shows starvation with J1. Note,
   BBR's unfairness in J1 is different from RTT-unfairness in traditional
   CCAs~\cite{reno,cubic}. For BBR, a small difference in RTprops leads to
   large unfairness that increases with the link rate~\cite{starvation}.

\anup{Tput ratios increase with capacity and this is different from RTT unfairness.}

\autoref{fig:kernel-validation} (Right) shows unfairness on parking lot with
5 ms RTprop. Copa matches the trend estimated by the contract. With BBR, the
  shape (derivative) of the contract is same as Copa. However the shift and
  RTT-bias in BBR's contract causes worse unfairness. For Cubic, the throughput
  ratio should be at least $\hops^{4/3}$ (contract is $\rate =
  \texttt{loss rate}^{-0.75}$~\cite{cubic-loss-model}). Reality is worse due
  to RTT-bias.

\section{Working around the tradeoffs}
\label{sec:workarounds}


As discussed in \autoref{sec:motivation}, we believe the only way to work around
the tradeoffs is to pick the input/output of the contract in a way that
decouples physical quantities (\eg, rate or delay) from the contract. We show
this for the four metrics.

Note that compound contract functions that take different shapes on different
link rates or switch the shape on the fly do not alleviate the tradeoffs. In
the worst-case, all the scenarios may occur simultaneously, \eg, multiple flows
per hop on a parking lot topology with noise.

\begin{packeditemize}

    \item {\bf Fairness.} As mentioned in \autoref{sec:metrics-tradeoffs},
      statistics that accumulate using max or min, like max per-hop delay,
      decouple multi-bottleneck fairness, trivially ensuring max-min fairness.
      However, such accumulation often relies on in-network
      support~\cite{poseidon}.

    \item {\bf Congestion and robustness.} The congestion growth metric
      describes growth in the statistic and not congestion. Decoupling
      statistic and congestion allows independently bounding congestion. For
      instance, \cite{ecn-vs-delay} shows use of a PI controller to obtain
      different ECN marking probabilities for the same queue buildup (\ie,
      different statistic for the same congestion). Likewise, explicit
      communication in packet headers (using enough precision) may eliminate
      noise to meet robustness~\cite{xcp}.

    \item {\bf Generality.} The domain of the statistic limits generality.
      Existing CCAs encode fair shares using a ``unary'' encoding. As proposed
      in \cite{starvation}, we can improve encoding efficiency using a
      ``binary'' encoding that communicates fair shares over time---similar to
      deriving multi-bit feedback from single ECN bit~\cite{abc}.
      Another work around is coordinate the ``fraction of link use'' (\ie, a
      quantity between 0 and 1) instead of ``absolute fair shares'' (\ie, an
      arbitrarily large number). This reduces the range of output values that
      a contract needs to support. BBR's rate-limited mode does
      this~\cite{bbr-rate-contract}, but BBR often operates in cwnd-limited
      mode~\cite{ray-bbr}, without fully leveraging this workaround.


\end{packeditemize}

\section{Limitations and future work}
\label{sec:limitations}

\myparagraph{Improving expressivity} In defining contracts, we faced a tradeoff
between expressivity (breadth of statements we can make) and tractability
(mathematically backing the statements). We erred on the side of tractability
to mathematically derive tradeoffs in \autoref{app:tradeoffs}. For instance,
our current definition makes assumptions about the network and other flows. As
a result, we are unable to reason about inter-CCA fairness. Due to similar
reasons, we also found it hard to prove/disprove that contracts are necessary
or sufficient for fairness.
Below, we describe the challenges, benefits, and possible approaches to improve
expressivity.



\mysubparagraph{Challenges} Two CCAs may achieve fairness even when they have
different contracts.
Consider Copa, it employs a delay-based contract when competing with itself,
but it switches to emulating Reno when it detects Reno flows. Even CCAs like
Vegas, that employ a delay-based contract all the time, may compete fairly with
Reno depending on the network conditions~\cite{num-low}. For instance,
RED-based packet drops~\cite{red} create a mapping between queuing delay and
loss rates. If the loss-rate-based and delay-based fair shares match, then
Vegas and Reno may compete fairly.




\mysubparagraph{Benefits} Improving expressivity would further guide CCA
design. For instance, to meet TCP-friendliness, BBRv3~\cite{bbr-ietf} leaves
``headroom'' for loss-based CCAs. If contracts are necessary for fairness, then
it is better to explicitly follow Reno's contract on detecting competing Reno
flows than leaving headroom which may or may not cause BBRv3 to follow Reno's
contract.

\mysubparagraph{Approach} We hope to extend reasoning using the formal methods
literature that also uses contract-like abstractions to reason about
distributed algorithms. For instance, \cite{contract-synthesis} defines
contracts as a \emph{set} of traces described using $\omega-$regular grammars,
while we defined contracts as a \emph{function}.

\myparagraph{Extending blueprints and applications of contracts} Our blueprints do not
cover all scenarios and metrics that CCAs may care about, \eg, short flows,
incast, \etc Performance in these scenarios often depends on other design
choices, \eg, slow start, receiver-based or credit-based control, \etc We hope
to convert our blueprints into a living repository to cover these evolving
scenarios.

Outside of designing and analyzing CCAs, we believe a contracts-first approach
would also improve reverse-engineering and classification of CCAs in the
wild~\cite{inspector-gadget, abagnale, ccanalyzer, nebby}.

\smallskip\noindent\textbf{Measurement} of workloads and network environments
can guide which metrics to prioritize in the tradeoff space. For instance,
assessing fairness requires understanding how often flows experience multi-hop
congestion and the typical number of flows or hops involved. Recent
work~\cite{stop-cca-fairness} suggests that contention may be rare on the
Internet, implying fairness may be less critical in some contexts. For
robustness, it would be useful to quantify the size and frequency of
non-congestive delays and losses. On the workload front, we want to understand
which network-level metrics correlate with application-level metrics. For
instance, \cite{sudarsanan-ml-cc} shows that unfairness is better for AI
collectives.

\section{Conclusion}
\label{sec:discussion}


We showed that contracts
determine key performance metrics, resulting in tradeoffs. We identify pitfalls
to avoid when designing CCAs. We hope that with our work, contracts will be a
conscientious design choice rather than an afterthought. Contracts should be a
direct consequence of desired steady-state performance, and rate updates should
be a consequence of desired reactivity/convergence time.




\bibliographystyle{ACM-Reference-Format}
\bibliography{references.bib}

\appendix

\section{Tradeoffs}
\label{app:tradeoffs}

We show that metrics (M1) robustness and (M2) fairness are at odds with (M3)
congestion and (M4) generality. M1 \& M2 require the contract to be gradual,
while M3 \& M4 require the contract to be steeper. We derive these tradeoffs
quantitatively. We assume the contract function \func has domain $[\Smin,
\Smax]$ and range $[\Cmin, \Cmax]$, \ie, $\Cmax = \func(\Smin)$, and $\Cmin =
\func(\Smax)$.

Our strategy to derive the tradeoffs is as follows, choice of a metric puts
constraints on the contract, and this in-turn puts constraints on other
metrics. So we fix one metric and see what constraints it puts on other
metrics.

The tradeoffs involving fairness depend on \agg. We consider $\agg \in \{\sum,
\max, \min\}$. The tradeoffs exists for $\sum$. We do not get tradeoffs for
$\max$ and $\min$, \ie, we can achieve max-min fairness independent of
requirements on robustness/generality. We also show derivation steps for
arbitrary \agg, if future work wants to consider other statistics that
accumulate differently across hops.


\subsection{M1 vs. M3: robustness vs. congestion growth}

Say we want the robustness error factor to be at most $\ef > 1$, then we
compute a lower bound on $\grow(n)$. From the definition of robustness error
factor (\autoref{eq:error-factor-def}):
\begin{alignat}{3}
    & \forall \s. & \frac{\func(\s)}{\func(\s + \ds)} & \leq \ef \nonumber\\
    & \text{Picking \s = \Smin, } \nonumber\\
    & \implies & \frac{\func(\Smin)}{\func(\Smin + \ds)} & \leq \ef \nonumber\\
    & \implies & \Cmax = \func(\Smin) & \leq \ef * \func(\Smin + \ds) \nonumber \\
    & & & \leq \ef * \ef * \func(\Smin + 2 * \ds)\nonumber\\
    & & & \leq \ef^{k} * \func(\Smin + k * \ds)\nonumber\\
    & \implies & \frac{\Cmax}{\ef^k} & \leq \func(\Smin + k * \ds) \label{eq:app:t1:tight}\\
    & \implies & \func^{-1}\left(\frac{\Cmax}{\ef^k}\right) & \geq \Smin + k *
    \ds \label{eq:app:t1:inverse}\\
    & \implies &
    \frac{\func^{-1}\left(\frac{\Cmax}{\ef^k}\right)}{\func^{-1}(\Cmax)} & \geq
    \frac{\Smin + k * \ds}{\Smin} \nonumber\\
    & \implies & \grow(\ef^k) \geq
    \frac{\func^{-1}\left(\frac{\Cmax}{\ef^k}\right)}{\func^{-1}(\Cmax)} & \geq
    \frac{\Smin + k * \ds}{\Smin} \label{eq:app:t1:growth}\\
    & \implies & \grow(n) \geq
    \frac{\func^{-1}\left(\frac{\Cmax}{n}\right)}{\func^{-1}(\Cmax)} & \geq
    \frac{\Smin + \frac{\log(n)}{\log(\ef)} * \ds}{\Smin} \label{eq:app:t1}
\end{alignat}

Note, we get \autoref{eq:app:t1:inverse} by taking $\func^{-1}$ on both sides,
the inequality flips as \func and $\func^{-1}$ are monotonically decreasing. We
get the left inequality in \autoref{eq:app:t1:growth} from the definition of
\grow (\autoref{eq:growth-def}). \autoref{eq:app:t1} shows that lower error
factor implies higher signal growth. The inequalities become equalities when
(by substituting $\Smin + k * \ds$ by \s in \autoref{eq:app:t1:tight}):
\begin{align}
\func(\s) = \Cmax\ef^{\frac{\s-\Smin}{\ds}}
\end{align}
This is same as the exponential contract from~\cite{starvation}.


\subsection{M1 vs. M4: robustness vs generality}

We compute an upper bound on $\frac{\Cmax}{\Cmin}$ given we want the error
factor to be at most $\ef > 1$. We start from \autoref{eq:app:t1:tight}, and
substitute $\Smin + k * \ds$ by $\Smax$:
\begin{alignat*}{3}
    & & \frac{\Cmax}{\ef^k} & \leq \func(\Smin + k * \ds)\\
    & \implies & \Cmax \ef^{-\frac{\Smax-\Smin}{\ds}} & \leq \func(\Smax) = \Cmin\\
    & \implies & \frac{\Cmax}{\Cmin} & \leq \ef^{\frac{\Smax-\Smin}{\ds}}
\end{alignat*}
Lower the robustness error, lower is the range of bandwidths we can support.
The inequality becomes equality for the exponential contract.



\subsection{M2 vs. M3: fairness vs. congestion growth}

We derive a lower bound on $\grow(n)$ given that we want the throughput ratio in
parking lot for $k$ hops to be $\ratiostar(k)$. For the parking lot ratio to be
$\ratiostar(k)$, we need:
\begin{align}
    & \forall k. \quad \max_{\s} \frac{\func(\s)}{\func(\agg(\s, \s, \ldots \s))} = \ratiostar(k) \nonumber
\end{align}
Say, the maximum of LHS is achieved when the link capacity in the parking lot
is $\bw = \bwstar(k)$. $\ratiostar(.)$ and $\bwstar(.)$ are functions of k. For
brevity, we drop the k, and refer to them as \rstar and \bwstar respectively.
The statements are true for all positive $k$.

Instead of directly computing a constraint on $\grow(n)$, we first derive a
constraint on $\grow_{\bwstar}(n)$. This eventually constrains $\grow(n)$.
Where, $\grow_{\bw}(n)$ is defined as
$\frac{\func^{-1}(\bw/n)}{\func^{-1}(\bw)}$. \Ie, $\grow(n) = \max_{\bw}
\grow_{\bw}(n)$ (from \autoref{eq:growth-def}). Note, $\grow_{\bw}(1) = 1$ from
this definition. We will use this later.

Consider the execution of the contract (CCA) on a parking lot topology with $k$
hops and link capacity $\bwstar$. In steady-state, we have:
\begin{align}
    \x[0] + \x[k] &= \bwstar \quad \text{same as \autoref{eq:fairness:balance}, and,} \nonumber\\
    \frac{\x[0]}{\x[k]} &= \rstar \quad \text{from the definition of \rstar above.} \nonumber
\end{align}
On solving these, we get:
\begin{align}
\x[0] = \bwstar / (\rstar + 1) \quad \text{and, } \quad \x[k] = \bwstar \rstar / (\rstar + 1) \nonumber
\end{align}
Say the aggregate statistic seen by flow $\f[i]$ in the parking lot is $\s[i] =
\func^{-1}(\x[i])$. Then $\grow_{\bwstar}(\bwstar/\x[i]) =
\frac{\func^{-1}(\bwstar/(\bwstar/\x[i]))}{\func^{-1}(\bwstar)} =
\frac{\s[i]}{\func^{-1}(\bwstar)}$. We define $\sstar = \func^{-1}(\bwstar)$,
and substitute $i$ by $0$. On rearranging, we get:
\begin{align}
    & \s[0] = \func^{-1}(\bwstar) \cdot \grow_{\bwstar}(\bwstar/\x[0]) = \sstar \cdot \grow_{\bwstar}(\rstar + 1) \nonumber\\
    \text{Likewise, } & \s[k] = \sstar \cdot \grow_{\bwstar}(1 + 1/\rstar) \nonumber
\end{align}
In steady-state:
\begin{align}
    & \s[0] = \agg(\s[k], \s[k], \dots, k \text{ times}) \quad \text{same as \autoref{eq:fairness:agg}, on substituting \s[0] and \s[k] just computed, we get} \nonumber\\
    \implies & \sstar \cdot \grow_{\bwstar}(\rstar + 1) = \agg (\sstar \cdot \grow_{\bwstar}(1 + 1/\rstar), \dots, k \text{ times})\nonumber
\end{align}
If multiplication distributes over accumulation, this is true for $\agg \in
\{\sum, \max, \min\}$, then:
\begin{align}
    \grow_{\bwstar}(\rstar + 1) = \agg (\grow_{\bwstar}(1 + 1/\rstar), \dots, k \text{ times})\nonumber
\end{align}
We consider under different choices of \agg, what constraint \ratiostar puts on
$\grow_{\bwstar}$. For \agg as $\max$ or $\min$. We can vacuously meet this
constraint for $\ratiostar = 1$ (corresponding to max-min fairness), and there
is no tradeoff. For $\agg = \sum$, the constraint becomes:
\begin{align}
    \grow_{\bwstar}(\rstar + 1) = k \cdot \grow_{\bwstar}(1 + 1/\rstar) \nonumber
\end{align}
Let $n = \rstar + 1 = \ratiostar(k)+1$, \ie, $k = \ratio^{\star-1}(n-1)$, where
$\ratio^{\star-1}$ is the inverse of $\ratiostar$ then:
\begin{align}
    \grow_{\bwstar}(n) = \ratio^{\star-1}(n-1) \cdot \grow_{\bwstar}(1 + 1/(n-1)) \nonumber
\end{align}

Since, \func is decreasing, $\grow_{\bw}$ is increasing for all \bw. So
$\grow_{\bwstar}(1 + 1/(n-1)) \geq \grow_{\bwstar}(1) = 1$. So, we get:
\begin{align}
    \grow_{\bwstar}(n) \geq \ratio^{\star-1}(n-1) \nonumber
\end{align}
If we want better fairness, we need \ratiostar to be slow growing, then
$\ratio^{\star-1}$ is fast-growing, and so $\grow_{\bwstar}$ is fast-growing,
and so congestion growth $\grow \geq \grow_{\bwstar}$ needs to be fast-growing.

For proportional fairness, $\ratiostar(k) = k$, and so $\grow(n) = O(n)$, this
is met by Vegas. For $\alpha$-fairness, $\ratiostar(k) = \sqrt[\alpha]k$, and
so $\grow(n) = O(n^{\alpha})$, this is met by the contract $\func(\s) =
\frac{1}{\x^{\alpha}}$.

\subsection{M2 vs. M4: fairness vs. generality}

We compute an upper bound on $\frac{\Cmax}{\Cmin}$ given we want the throughput
ratio in parking lot for k hops to be at most $\ratiostar(k)$. For the
throughput ratio to be at most $\ratiostar(k)$, we need:
\begin{align}
& \forall \s, k. \quad \frac{\func(\s)}{\func(\agg(\s, \s, \ldots \s))} \leq \ratiostar(k) \nonumber
\end{align}
The derivation beyond this depends on \agg. For $\agg \in \{\max, \min\}$, the
above constraint is vacuously true and there is not tradeoff. For $\agg =
\sum$, we need:
\begin{align}
& \forall s, k. \func(\s) \leq \ratiostar(k) \cdot \func(k * \s) \nonumber
\end{align}
Picking $\s = \Smin$, and $k = \frac{\Smax}{\Smin}$, we get:
\begin{align}
& \func(\Smin) \leq \ratiostar\left(\frac{\Smax}{\Smin}\right) \cdot \func(\Smax) \nonumber\\
\implies & \frac{\Cmax}{\Cmin} \leq \ratiostar\left(\frac{\Smax}{\Smin}\right) \nonumber
\end{align}
Better fairness implies smaller ratio $\ratiostar(.)$, implies smaller the
range of supported bandwidths.

If we want fairness to be at least as good as proportional fairness, we need
$\ratiostar(k) \leq k$, or:
\begin{align}
    & \forall s, k. \quad \func(\s) \leq \ratiostar(k) \cdot \func(k * \s) \leq k \cdot \func(k * \s) \nonumber
\end{align}
Substituting $k = \Smin/\s$, we get:
\begin{align}
    & \func(\s) \leq \frac{\Smin\Cmax}{\s} \nonumber
\end{align}
Equality occurs when $\func(\s) = \frac{\Smin\Cmax}{\s}$, which is the same
contract as Vegas.

Likewise, for $\alpha$-fairness, $\ratiostar(k) = \sqrt[\alpha]k$, and so
$\frac{\Cmax}{\Cmin} \leq \sqrt[\alpha]{\frac{\Smax}{\Smin}}$. The equality
occurs for the contract: $\func(\s) = \frac{1}{\x^{\alpha}}$.


\section{Validation}
\label{app:validation}

\begin{figure*}
    \centering
    \includegraphics[width=0.33\textwidth]{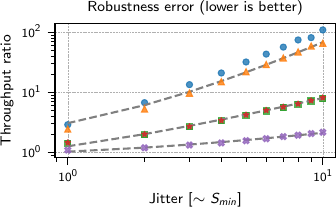}\hfill
    \includegraphics[width=0.33\textwidth]{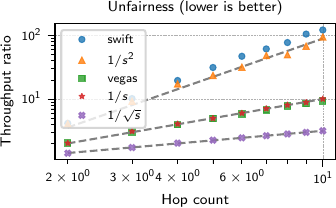}\hfill
    \includegraphics[width=0.33\textwidth]{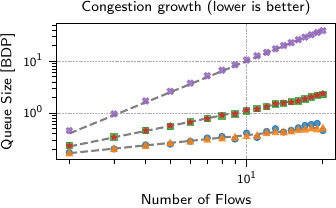}\hfill

    \caption{\label{fig:app:validation} Contracts-based performance metric
    estimates match packet-level simulation. The markers show empirical data
  and \textcolor{TextGray}{gray} lines show performance estimated by the
contracts.}

\end{figure*}

\autoref{fig:app:validation} shows the simulation results on a log-log scale
plot for visual clarity.

\end{document}